\documentclass[showkeys,aps,amssymb,twocolumn,prd]{revtex4}

\usepackage{bm,graphicx}

\begin{document}

%
\newcommand{\vect}[1]{\ensuremath{\mbox{\boldmath $#1$}}}
\def\bby  {{\bf y}}
\def\bc  {{\bf c}}
\def\beq{\begin{equation}}
\def\beqa{\begin{eqnarray}}
\def\beqan{\begin{eqnarray*}}
\def\bm {{\mathop{\rm Mag}\nolimits }}
\def\bmo  {{\bf 0}}
\def\bg {{\bf g}}
\def\br  { {\mbox{\boldmath$r$}} }
\def\brr  {{\sf r}}
\def\bs  { {\mbox{\boldmath$s$}} }
\def\bss  {{\sf s}}
\def\bt  {\vect{\theta}}
\def\bu  {{\bf u}}
\def\bv  {{\bf v}}
\def\bw  { {\mbox{\boldmath$w$}} }
\def\bx  {{\bf x}}
\def\bX  {{\bf X}}
\def\caustics {\mathop{\rm Caustics}\nolimits}
\def\cD { {\cal{D}} }
\def\cH { {\cal{H}} }
\def\cK { {\cal{K}} }
\def\cN { {{\cal{N}}^{\tau_0}_{\tau_S} (\br)}  }
\def\cL { {\cal{L}}  }
\def\cP { {\cal{P}} }
\def\crit {\mathop{\rm Crit}\nolimits}
\def\cQ { {\cal{Q}} }
\def\cS { {\cal{S}} }
\def\cT { {\cal{T}} }
\def\cW { {\cal{W}} }
\def\CC{{\bf C}}
\def\det {\mathop{\rm det}\nolimits}
\def\dls{d_{ls}}
\def\dol{d_{l}}
\def\dos{d_{s}}
\def\dsum  {\displaystyle\mathop{\sum}}
\def\eeq{\end{equation}}
\def\eeqa{\end{eqnarray}}
\def\eeqan{\end{eqnarray*}}
\def\grad {\mathop{\rm grad}\nolimits}
\def\Hess  {\mathop{\rm Hess}\nolimits}
\def\jac  {\mathop{\rm Jac}\nolimits}
\def\hbu{\hat{\bu}}
\def\hbt{\hat{\bt}}
\def\hp{\hat{\Psi}}
\def\hx  {\hat{x}}
\def\hy  {\hat{y}}
\def\kk  {{\mbox{\footnotesize \sf K}}}
\def\pa  {\partial}
\def\rank {\mathop{\rm rank}\nolimits}
\def\rd { {\rm d} }
\def\rK { {\rm K} }
\def\ru  {{\mbox{\rm u}}}
\def\rv  {{\mbox{\rm v}}}
\def\RR  {{\bf R}}
\def\sfg  {{\sf g}}
\def\sfL  {{\sf L}}
\def\sfr  {{\sf r}}
\def\sfs  {{\sf s}}
\def\sfS  {{\sf S}}
\def\sft  {{\sf t}}
\def\sfx  {{\sf x}}
\def\sign {\mathop{\rm sign}\nolimits}
\def\th {\theta}
\def\ttL  {{\tt L}}
\def\ttr  {{\tt r}}
\def\ttR  {{\tt R}}
\def\ttp  {{\tt p}}
\def\ttq  {{\tt q}}
\def\tts  {{\tt s}}
\def\ttt { {\tt t} }
\def\ttx { {\tt x} }
\def\tty { {\tt y} }
\def\ttz { {\tt z} }
\def\ttT  {{\tt T}}
\def\tx  {\tilde{x}}
\def\ty  {\tilde{y}}

\newcommand{\bme}{ {\mbox{\boldmath $\eta$}} }
\newcommand{\bma}{ {\mbox{\boldmath $\alpha$}} }


\title{\hfill\rightline{\small{\it }}
Wavefronts, Caustic Sheets, and Caustic Surfing in Gravitational Lensing}

\author{Simonetta Frittelli}
\affiliation{Department of Physics, Duquesne University,
Pittsburgh, PA 15282, USA}
\author{A.  O. Petters}
\affiliation{Department of Mathematics, Duke University,
Science Drive, Durham, NC 27708-0320, USA}
\hfill{To appear in the Journal of  Mathematical  Physics.}

\begin{abstract}

Very little attention has been paid to the properties of optical
wavefronts and caustic surfaces due to gravitational lensing. Yet
the wavefront-based point of view is natural and provides
insights into the nature of the caustic surfaces on a
gravitationally lensed lightcone. We derive analytically the basic
equations governing the wavefronts, lightcones, caustics on
wavefronts, and caustic surfaces on lightcones in the context of
weak-field, thin-screen gravitational lensing. These equations are
all related to the potential of the lens. In the process, we also
show that the standard single-plane gravitational lensing map
extends to a new mapping, which we call a wavefront lensing map.
Unlike the standard lensing map, the Jacobian matrix of a
wavefront lensing map is  not symmetric. Our formulas are then
applied to caustic ``surfing.'' By surfing a caustic surface, a
space-borne telescope can be fixed on a gravitationally lensed
source to obtain an observation of the source at very high
magnification over an extended time period, revealing structure
about the source that could not otherwise be resolved.  Using
our analytical expressions for caustic sheets, we present a scheme
for surfing a 
caustic sheet 
of a lensed source in rectilinear
motion. Detailed illustrations are also presented of the possible
types of wavefronts and caustic sheets due to nonsingular and
singular elliptical potentials, and singular isothermal spheres,
including an example of 
caustic surfing  for a
singular elliptical potential lens. 
\end{abstract}


\maketitle

\section{Introduction}

Among relativistic concepts of direct application to
gravitational lensing, the observer's past lightcone is perhaps
the most fundamental.
The lightcone concept unifies the temporal and spatial properties of
lensing events in a geometrical manner that makes the multiplicity,
magnification, and time delay  of the images arise naturally.  In this view,
gravitational lensing by dark and luminous matter causes the
observer's past lightcone to curve into a singular
three-dimensional hypersurface that self-intersects and folds
sharply. It is the singular ``folding'' of the lightcone that is
responsible for most features of interest in gravitational
lensing, such as image multiplicity, magnification, and time delay.
Additionally, in this view, caustics --- so relevant to
observational lensing --- literally acquire a new dimension,
turning into caustic sheets (possibly multiple)
contained within the lightcone itself, and carrying a sense of
time as well as spatial location.  The lightcone concept is most
naturally built by imagining an optical wavefront emanating from
an observer and traveling out and towards the past or future.  At the start,
the wavefront is convex and almost perfectly spherical. Upon
encountering gravitational lenses,  the wavefront becomes
non-uniformly retarded, acquiring dents. No matter how
small and insignificant these dents in the wavefront are soon
after leaving the neighborhood of the lens, in time the normal
propagation enhances them, inevitably developing crossings and
sharp ridges. The surface traced out in spacetime by a wavefront
is a (past or future) lightcone.  The trace of the wavefront's
sharp ridges and ``swallowtail'' points is called a ``comoving caustic surface'' 
if thought of as embedded in
three-space, and a ``caustic surface'' if viewed in spacetime.
In other words, a caustic surface is built out of caustic sheets with
cusp ridges and higher-order singularities (e.g., swallowtails,
elliptic umbilics, and hyperbolic umbilics).
Remarkably, in spite of the naturalness of this concept, very
little attention has been paid to the greater significance of the
caustics as sheets or surfaces.
As a result, to our knowledge, the
properties of optical wavefronts and caustic surfaces in
gravitational lensing have not progressed much beyond the generic
local classification of caustic singularities of the general theory as
developed by Thom, Arnold, and others. Our goal is to initiate a program  that
investigates optical wavefronts and caustics 
as they relate to the matter distribution of the
gravitational lenses in the spacetime.

The vast majority of gravitational lenses can be modeled extremely
well using a small-angle approximation of weak-field perturbations
of a Friedmann-Lema\^{i}tre universe, and projecting the deflectors
into planes (thin-screen approximation)  --- see the monographs by
Schneider, Ehlers,
and Falco \cite{SEF}  and Petters, Levine, and Wambsganss \cite{PLW} for a
detailed treatment. An
enormous advantage of weak-field, thin-screen gravitational lensing
is that we can obtain explicit analytical expressions for the
associated lightcones, optical wavefronts, and caustics on both the
wavefronts and lightcones. Indeed, one of the reasons there has
been limited progress in the study of quantitative 
properties of the
optical caustic surfaces and wavefronts due to gravitating matter
is the lack of computable physical models with observational
relevance. Thin-screen, weak-field gravitational lensing provides
us with such a setting.

Gravitational lensing from a wavefront perspective was used by
Refsdal \cite{R64} to relate the Hubble constant to the time delay
between lensed images (see \cite{R64}, Kayser and Refsdal
\cite{KR,B} for more). Arnold's singularity theory was used by Petters
\cite{Ptt93}  to treat the local qualitative properties of caustic
surfaces (big caustics) in gravitational lensing. Friedrich and
Stewart \cite{FS}, Stewart \cite{Stw90}, 
Hasse, Kriele and Perlick \cite{HKP96}, Low \cite{L97}, and
Ehlers and Newman \cite{EN00} also used Arnold's theory to treat
wavefronts and caustics in general relativity with varied
motivations. In the case of a strong gravitational field, Rauch
and Blandford \cite{RB94} numerically computed the caustic sheets due to the
Kerr metric. Wavefronts in the context of liquid droplet lenses
have been studied extensively by Berry, Hannay, Nye, Upstill, and
others --- see Berry and Upstill \cite{BU80}, Nye \cite{Ny99}, and
references therein. Readers may also benefit from the popular
article by Nityananda \cite{Nt90} on wavefronts in gravitational
lensing.

Our paper calculates explicitly the equations governing the
wavefronts and caustic surfaces in gravitational lensing with the
standard approximations, and analytically relates the wavefronts
and caustic surfaces to the gravitational potential of the lens.
This allows us to give a first quantitative treatment of {\it
caustic surfing,\/} a futuristic notion suggested by Blandford \cite{Bl01}
in his millennium essay. The motivation for caustic surfing
is to lock a satellite on a gravitationally lensed moving source so
as to observe the source at a high magnification over an extended
time period. We derive the starting equations 
for the trajectory to
be followed in order to surf the caustic surface.  The appeal of
caustic surfing lies in its potential as a tool to obtain
information about a distant moving source that could not otherwise
be resolved.

Our article is organized as follows.  Section II derives an
explicit expression for the present optical path length in units of time of light signals that reach
the observer in the thin-screen approximation. In Section III, this
expression is used to define a new lensing map in which the sources
lie on an instantaneous wavefront, rather than on a lens plane. The
caustic surfaces and sheets are related to the wavefront lensing
map in Section IV. Section V develops some fundamentals of the
concept of caustic surfing. Applications to the case of nonsingular and
singular elliptical potential models, as well of singular
isothermal spheres, are given in Section VI.  The latter section also gives
full
classifications of the singularities of the wavefronts and caustic surfaces  of
the noted lenses, and
explicitly illustrates the
caustic surfing concept.  We conclude in Section VII with some
remarks and an outlook of future applications. The research
reported in this paper extends preliminary results on wavefronts in
gravitational lensing first reported by Frittelli and Petters
\cite{FP01} at the Ninth Marcel Grossmann Meeting.

\section{The Present Proper Length Function and Lens Equation}

We shall calculate the present optical length of gravitationally lensed
light rays from a source to the observer using Fermat's
principle and coordinates in space. This will include a formula
(lensing map) relating the position of a lensed source to the
impact positions at the gravitational lens of lensed light rays
from the source to observer.
For the convenience of readers who are new to gravitational
lensing, Sections II.A  and II.B.1 give an introduction overviewing
those aspects of lensing relevant to our wavefront and caustic surface
considerations.  Readers already familiar with the basics of gravitational
lensing can skip to Section II.B.2, where we cast the lens equation
in comoving form.

\subsection{Cosmological Model for Lensing}

We use the usual cosmological setting of gravitational lensing,
namely, a weak-field perturbation of a Friedmann-Lema\^{i}tre
spacetime.  These models are quite robust and fit extremely
well with observations.  A detailed treatment of the
assumptions and approximations  used to model most
gravitational lensing scenarios is given in Chapter 3 of
\cite{PLW}.

The spacetime geometry   in the vicinity of a weak-field
gravitational lens system  (i.e., source, lens(es), observer,
and light rays between source and observer) is  approximated by
the following spacetime metric:

\beqa
\label{eq-GLmetric}
        \bg_{GL}^\kk &=&
                -\left (1+2\frac{\phi}{c^2}\right )
                 c^2 d \tau^2 +
                a^2(\tau)
                \left (1-2\frac{\phi}{c^2}\right ) d S^2_\kk\nonumber \\
        &=& a^2(\ttt)\left [-\left (1+2\frac{\phi}{c^2}\right )
                 d \ttt^2 +
        \left (1-2\frac{\phi}{c^2}\right )d S^2_\kk\right]. \nonumber \\
\eeqa

\noindent The quantity $\tau$ is cosmic time, $\ttt$ is conformal
time related to $\tau$ by

\beq
\label{eq-timerates}
 d \ttt = \frac{c}{a (\tau)} d \tau,
\eeq

\noindent with $a (\tau)$ a positive function called the {\it
expansion factor} and we use the notation $a(\ttt) = a(\tau
(\ttt))$. Also $\phi$ represents the time-independent Newtonian
potential of the density perturbation due to the lens(es), and $d
S^2_\kk$ the metric of space with constant curvature $\kk =
-1,0,1$.  The weak-field limit is assumed (i.e., $|\phi|/c^2 \ll
1$), so we ignore terms of order greater than $\phi/c^2$.  Note
that the potential $\phi$ obeys the cosmological Poisson equation
--- see Section~\ref{subsub-T-le}. The
weak-field assumption near a gravitational lens ensures that the
bending angles $\hat{\alpha}$ of light rays deflected by the lens
are small, i.e., $\hat{\alpha} \ll 1$. In addition to the latter
assumption,  we  suppose that the lens is ``thin.'' This means
that along the line of sight, the diameter of the matter
distribution of the lens or source  is very small compared to the
distances between lens and observer, and between lens and source.
We shall treat the gravitational lens and the point sources as
lying on planes --- called the {\it lens plane} and {\it light
source plane} --- approximately orthogonal to an axis defined by
the observer and some central point on the lens -- referred to as
the optical axis. All these assumptions allow us to consistently
restrict lensing to small observation angles as measured from the
optical axis.

Now consider the gravitational lensing metric $\bg_{GL}^\kk$ in
(\ref{eq-GLmetric}). There are coordinates $(R, \theta,
\varphi)$ in space, called {\it comoving coordinates},  such
that  the Riemannian metric $dS^2_\kk$ becomes

\beq
\label{eq-ccmetric}
dS^2_\kk =
      \frac{dR^2}{1-\kk R^2} +
 R^2 (d\theta^2 + \sin^2\theta d\varphi^2).
\eeq

\noindent We assume that the comoving coordinates are
dimensionless, the expansion factor $a (\tau)$ has physical
dimensions of length, and the potential $\phi$ has physical
dimensions of $\mbox{\rm [length]}^2/\mbox{\rm [time]}^2$. The
cosmic time $\tau$ has dimension of time, while  the conformal
time $\ttt$ is dimensionless. Away from the neighborhood of the
lens, the gravitational lensing metric $\bg_{GL}^\kk$ is
approximated by the Friedmann-Lema\^{i}tre (FL) metric

$$\bg^\kk_{\rm FL}
=
- c^2 d \tau^2 +
 a^2(\tau) \
d S^2_\kk = a^2(\ttt) \left[
                 - d \ttt^2 +
        d S^2_\kk\right].
$$

\noindent This is employs the assumption that the Newtonian
potential $\phi$ decreases to zero sufficiently fast
away from the lens with the lens-source and observer-lens
angular diameter distances sufficiently large.
In a realistic physical sense, the
Newtonian potential of an isolated body decays no faster than
$r^{-1}$.   Some plausible ways to justify the use of this metric as
intended are discussed in \cite{EFN01} and references therein. Still,
it is not at all clear how to improve on this assumption rigorously.
Indeed, there are some
unresolved technical mathematical issues with the standard approximations in
gravitational lensing, but it is not our intention to
address them here.

In this work we will restrict our discussion to a flat
cosmological setting. {\it Henceforth, we shall assume that the
gravitational lensing metric $\bg^\kk_{GL}$ has $\kk =0$.} The
associated FL metric then reduces to the Einstein-de Sitter metric

\beq
\label{eq-FL-flat}
\bg_{\rm Ed} \equiv \bg^0_{\rm FL}
=
a^2(\ttt)\  ds^2_{\rm flat},
\eeq

\noindent where

\beq
\label{metricflat}
ds^2_{\rm flat}
=
 - d \ttt^2 +  dS^2_0, \qquad 
dS^2_0 =   dR^2  +
 R^2 (d\theta^2 + \sin^2\theta d\varphi^2)
\eeq

\noindent with $0 \le R < \infty,$ \ $0 \le \theta < \pi,$  \ $0
\le \varphi < 2 \pi$. Note that for $\kk =0$ the comoving
coordinates $(R, \theta, \varphi)$  define spherical coordinates
in space. The conformal relationship (\ref{eq-FL-flat}) yields
that the Minkowski metric $ds^2_{\rm flat}$ is very important for
our study,  because $\bg_{\rm Ed}$ and $ds^2_{\rm flat}$ have the
same null geodesics up to an affine parameter.  The topology of
our spacetime with a gravitational lens now takes the form $I
\times E$, where $I$ is an open interval of $\RR$ and $E$ an open
subset of $\RR^3$. In most applications of gravitational lensing,
we have  $E = \RR^3 - A$, where $A$ is a finite set of points
corresponding to singularities (e.g., point masses) in the lens.
These singularities would include points where the density
perturbation of the lens or the Newtonian potential $\phi$
diverge.  This yields a lens plane $L$ that is $\RR^2$ minus a
finite set of points. We call $E$ the {\it comoving space} and
$\tau \times E$ the {\it proper three-space} at cosmic time
$\tau$. Away from the lens, the metrics on $E$ and $\tau \times E$
are the standard Euclidean metric $dS^2_0$ and metric  $a^2 (\tau)
\ dS^2_0$, respectively. When we study the caustic surfaces along
our past lightcone, we project them into $E$.

\subsection{The Present Proper Length Function and the Comoving Lens Equation}

\subsubsection{Physical Setting for Present Proper Length of Light Rays}

We must discuss the basic physical setting needed to determine the
present optical length of light rays via Fermat's principle (see pp. 65-76
of \cite{PLW} for a detailed treatment). Suppose that a
light signal emanates from a source at cosmic time $\tau_S$ and
arrives at the observer at the present cosmic time $\tau_0$.
Assume that the observer is at the origin $\bmo$ of the spherical
coordinate system $(R, \theta, \varphi)$ in the comoving space $S$
and the source at comoving position $p_S = (R_S, \theta_S,
\varphi_S)$. Assume that the light ray undergoes deflection by a
gravitational lens on some intermediate lens plane $L$. Since the
deflection angle at the lens plane is assumed small, we
approximate the light ray's trajectory by an FL piecewise-smooth
null geodesic consisting of an FL null geodesic from the source to
$L$ and one from $L$ to the observer.   Fix an optical axis
through  an arbitrary comoving point $p_L$ on $L$ (e.g., center of mass
of the lens); say, $p_L = (R_L, \theta_L, \varphi_L)$ with $(R,
\theta_L, \varphi_L)$  defining the optical axis. Take $p_L$ to be
the origin of $L$.   We now consider the set $\cal N$ of  all
piecewise-smooth null FL geodesics 
that left the source at cosmic
time $\tau_S$ and arrive at the observer at roughly  the present
cosmic time $\tau_0$ after undergoing deflection at $L$.
The
travel time differences between these null rays are assumed to be
significantly smaller than  the Hubble time ($1/H_0$) and so the
scale factor is assumed to change negligibly during these 
time differences.  In particular, the scale
factor is taken to be approximately  $a(\tau_0)$ when the
different rays arrive. Fixing $\tau_S$ and $\tau_0$, the set $\cN$
of null rays can be parametrized by their comoving impact vectors
$\br$ on $L$ (relative to the origin of $L$). The rays in $\cN$
will be denoted by $\nu_\br$.

But not all the rays in $\cN$ represent actual light rays.
By Fermat's
principle (e.g., pp. 66-67, \cite{PLW}), the actual
physical light rays (within our approximations) are given by the
rays in $\cN$  whose present proper lengths are stationary with respect to
variations of the comoving impact vector $\br$. To characterize the
light rays, we first determine a formula for the present proper 
lengths of
the rays in $\cN$. Our spacetime with
\beq
\label{eq-GLmetric-0}
\bg_{\rm GL} \equiv \bg^0_{\rm GL}
=
 a^2(\ttt)\left [-\left (1+2\frac{\phi}{c^2}\right )
                 d \ttt^2 +
        \left (1-2\frac{\phi}{c^2}\right )d S^2_0\right]
\eeq
can be viewed as an Einstein-de Sitter universe with a
medium having the following time-independent refractive index
(e.g., p. 54, \cite{PLW}):

\beq
\label{eq-refractive-index}
n   = 1  \ - \frac{2}{c^2} \ \phi.
\eeq

\noindent Along a null ray  $\nu_\br (s) = (\tau (s),
\hat{\nu}_\br (s))$ in $\cN$, where $s$ is a parameter along the
ray and $\hat{\nu}_\br (s)$ is the
spatial path of  $\nu_\br$ in $E$, equation
(\ref{eq-GLmetric-0}) shows that

\beq
\label{eq-refindex-ray}
n(\nu_\br(s))  = 1  \ - \frac{2}{c^2} \ \phi (\hat{\nu}_\br (s))
 = \left| \frac{d \ttt}{d \ell} \right|,
\eeq

\noindent where  the dimensionless incremental length $d \ell$
along $\hat{\nu}_\br (s)$ is relative to the Euclidean metric $d
S^2_0.$ The rightmost equality in equation (\ref{eq-refindex-ray})
uses the assumption that $|\phi|/c^2 \ll 1$ and terms of order
higher than $\phi/c^2$ are ignored.

The present proper length $L_{\tau_0} (\nu_\br)$  of a ray
$\nu_\br$ in $\cN$ is then the ray's optical length  relative to
the refractive medium in the hypersurface $\tau_0 \times E$. In other words,

$$ L_{\tau_0} (\nu_\br)
= \int_{\hat{\nu}_\br} \ n(\hat{\nu}_\br) \  d \ell_{\tau_0}.$$

\noindent Here the increment of length $d \ell_{\tau_0}$ is
with respect to the  spatial metric   $a_0^2 \ d S_0^2$ on
$\tau_0 \times E$, where

$$a_0 \equiv a (\ttt (\tau_0)).$$

\noindent Note that $d \ell_{\tau_0} = a_0 \ d \ell$. Using
(\ref{eq-refractive-index}), the present proper length of $\nu_\br$ 
in units of time is

\beq
\label{eq-travel-time}
\cT (\nu_\br)  = \frac{L_{\tau_0}(\nu_\br)}{c}
= \frac{a_0}{c} \ \int_{\hat{\nu}_\br} \left(1  \ -
               \frac{2}{c^2} \ \phi(\hat{\nu}_\br) \right)\ d \ell.
\eeq
{\it Henceforth, we shall refer to $\cT$ simply as a 
present proper length function and assume that its values are in units of time.}

By Fermat's principle, the light rays are characterized by those
rays in $\nu_\br$ for which $\cT (\nu_\br)$ is a stationary value
of $\cT$. In the next subsection, we shall obtain an explicit
formula that determines these stationary values.

\subsubsection{Present Proper Length Function and Lens Equation
via Comoving Coordinates}
\label{subsub-T-le}

We shall now provide a practical expression of the present proper length (in units of time) 
in terms of rectangular coordinates in the comoving space
$E$ of our lensing spacetime $(I\times E, \bg_{GL})$. First, let's
transform the dimensionless spherical coordinates $(R, \theta,
\varphi)$ to dimensionless rectangular coordinates $(x,y,z)$ such
that the observer is at $(0,0,0)$ and  the optical axis $(R,
\theta_L, \varphi_L)$ corresponds to the $z$-axis.   The lens
plane $L$ is the $xy$-plane located at $z_l$ along the positive
$z$-axis.  We shall re-label the rectangular coordinates on $L$ as
$(\br,z_l)$, where $\br = (r_1, r_2)$ and $z_l>0$ is fixed. The
source is assumed to be located at $(\bs, z)$, where $\bs = (s_1,
s_2)$. When $z > z_l$, the source is behind the lens (and so is
gravitationally lensed).

The cosmological Poisson equation at the present cosmic time $\tau
= \tau_0$ is \beq \label{eq-cosmo-pe} \nabla^2_\sfx \  \phi (\sfx)
= 4 \pi G a^2_0 \  \rho (\sfx), \eeq where $\sfx = (x,y,z)$,
$\nabla^2_\sfx$ is the Laplace operator relative to $\sfx$, $G$ is
the gravitational constant, and $\rho$ is the cosmological volume
mass density.  Here $\rho$ has units of ${\rm [mass]/[length]^3}$.
Set
$$\tilde{\rho} \equiv a^3_0 \rho, \qquad \tilde{\phi} \equiv a_0 \phi.$$
Since (\ref{eq-cosmo-pe}) is solved by
$$\phi (\sfx) = -a^2_0 G \int_{\RR^3} \rho (\sfx') \frac{d \sfx'}{|\sfx - \sfx'|},$$
we have
$$\tilde{\phi} (\sfx) = - G \int_{\RR^3} \tilde{\rho} (\sfx')
\frac{d \sfx'}{|\sfx - \sfx'|},$$
which has units of ${\rm [length]^3/[time]^2}$.
Integrating (\ref{eq-cosmo-pe}) along the optical axis from
the source to the observer, we obtain the cosmological 2D-Poisson
equation:
\beq
\label{eq-cosmo-pe-2D}
\nabla^2_\br \Psi (\br) = \frac{8 \pi G a^2_0}{c^2} \ \sigma (\br),
\eeq
where
$$\Psi (\br) \equiv \frac{2}{c^2} \int^0_z \phi (\br, z') a_0 d z',
\qquad \sigma (\br) \equiv \int^0_z \rho (\br, z') a_0 d z'.$$
Note that $\Psi (\br)$ has units of length, while
$\sigma (\br)$ has units of ${\rm [mass]/[length]^2}$.
Let
$$
\tilde{\Psi} (\br) = \frac{2}{c^2} \int_z^0 \tilde{\phi}(\br, z') d z',
\qquad
\tilde{\sigma} (\br) = \int_z^0 \tilde{\rho}(\br, z') d z'.
$$
It follows that $\tilde{\Psi} = \Psi$ and $\tilde{\sigma} = a^2_0
\sigma.$ Equation (\ref{eq-cosmo-pe-2D}) then yields the comoving
2D-Poisson equation: \beq \label{eq-comov-pe-2D} \nabla^2_\br
\tilde{\Psi} (\br) = \frac{8 \pi G }{c^2}\  \tilde{\sigma} (\br),
\eeq where $\tilde{\Psi}$ and $\tilde{\sigma}$ have units of
length and mass, respectively. The function $\tilde{\Psi}$ can be
expressed formally as a solution of (\ref{eq-comov-pe-2D}) by
$$
\tilde{\Psi} (\br)
        = \frac{4G}{c^2}
        \int_{\RR^2} d \br'\tilde{\sigma} (\br') \
                \ln \left| \frac{\br'-\br}
                                {\xi_0}\right|,
$$
where $\xi_0$ is an arbitrary, dimensionless, fixed constant.

Equation (\ref{eq-travel-time}) yields that  the present proper length of
a null ray $\nu_\br$ in $\cN$ separates into two terms:
\beq
\label{eq-cT-gen}
\cT (\nu_\br) = \frac{a_0}{c} \left[ L (\hat{\nu}_\br) \ -
                    \ \hat{\Psi} (\br)\right],
\eeq
where
$$L (\hat{\nu}_\br) \equiv \int_{\hat{\nu}_\br} \ d \ell,
\qquad \quad
\hp (\br) \equiv \frac{2}{c^2} \ \int_{\hat{\nu}_\br} \
                          \phi (\hat{\nu}_\br) \ d \ell
$$
with  $L (\hat{\nu}_\br)$ the (dimensionless) Euclidean length of
$\hat{\nu}_\br$.  Since bending angles are assumed small, we approximate
the line integral of $\phi$ over $\hat{\nu}_\br$ by an integral along
the optical axis, i.e.,
$$\hp (\br) = \frac{2}{c^2} \ \int_z^0 \
                          \phi (\br, z') d z'
= \frac{\tilde{\Psi} (\br)}{a_0}.
$$
Note that the scaled comoving potential $\hp (\br)$ is dimensionless.

The contribution of the term $a_0 L
(\hat{\nu}_\br)/c$ to equation (\ref{eq-cT-gen})
is  the time due to the present proper length of the light's
geometric path relative to the Euclidean metric, while  the
term $a_0 \hat{\Psi} (\br)/c$ contributes the relativistic time
dilation (Shapiro time delay) due to the gravitational potential $\phi$ 
of the lens.
We have

\beq \label{eq-L-ells}
L (\hat{\nu}_\br) = \ell_l \  +  \ell_{ls},
\eeq

\noindent where $\ell_l$ is the Euclidean length of the segment
of $\hat{\nu}_\br$ consisting of the straight line from
comoving impact vector $\br$ to the observer and $\ell_{ls}$
the Euclidean length of the straight-line segment  from the
source to $\br$.  Expressing the null ray $\nu_\br$
parametrically as $\nu_\br (s) = (\tau(s), \hat{\nu}_\br (s))$,
we obtain

$$
0 = \bg_{\rm Ed} \left(\frac{d\hat{\nu}_\br}{ds}, \frac{d \hat{\nu}_\br}{ds}\right)
= -c^2 \left(\frac{d\tau}{ds}\right)^2 \ + \ a^2 (\tau)
  \left|\frac{d\hat{\nu}_\br}{ds}\right|^2,
$$

\noindent where the squared magnitude of the comoving velocity
$d \hat{\nu}_\br/ds$ is  relative to $d S^2_0$. This implies

\beqan
\ell_l & = & \int_{s_L}^{s_0} \left|\frac{d\hat{\nu}_\br}{ds}\right| \ ds
=  \int_{\tau_L}^{\tau_0} \frac{c}{a (\tau)} \frac{d\tau}{ds} \ ds \\
& = & \int_{\ttt_L}^{\ttt_0} d \ttt = \ttt_0 \ - \ \ttt_L,
\eeqan

\noindent where $\tau_L$ is the cosmic time when $\nu_\br$
arrives at the lens plane, $s_0$ and $s_L$ are the respective
parameter values  corresponding to $\tau_0$ and $\tau_L$, and
$\ttt_0 = \ttt (\tau_0)$, $\ttt_L = \ttt (\tau_L)$. Similarly,

$$\ell_{ls} = \ttt_L - \ttt_S,$$

\noindent where $\ttt_S = \ttt (\tau_S)$ with $\tau_S$ the
cosmic time when $\nu_\br$ left the source. Hence, the present proper length
function $\cT (\nu_\br)$ can be expressed in terms of conformal
time as

\beq
\label{eq-c-cT}
\cT (\nu_\br) = \frac{a_0}{c} \ \left[(\ttt_0 \ - \ \ttt_S) \ - \ \hat{\Psi} (\br)\right].
\eeq

We now express the present proper length function in terms of our comoving
rectangular coordinates. In the
approximation of small angles, we have

\beqa
\label{eq-ells}
\ell_l &=& \sqrt{z^2_l + |\br|^2} = z_l + \frac{|\br|^2}{2z_l}
\nonumber\\
\ell_{ls}&=&  \sqrt{(z-z_l)^2 + |\br - \bs|^2}
          = z-z_l +
        \frac{|\br - \bs|^2}
             {2(z-z_l)}.
\nonumber\\
\eeqa

\noindent By (\ref{eq-L-ells}) and (\ref{eq-ells}), the present proper
length of the null ray 
$\nu_\br$ becomes

\beq
\label{eq-cT}
    \cT (\br, \bs, z) \equiv \cT (\nu_\br)
    =
    \frac{a_0}{c^2} \ttT (\br, \bs, z),
\eeq

\noindent where

\beq
\label{eq-cT-1}
\ttT (\br, \bs, z)
\equiv z + \frac{|\br|^2}{2z_l}
      + \frac{|\br - \bs|^2}{2(z-z_l)}
       -\hat{\Psi}(\br).
\eeq

\noindent Equation (\ref{eq-c-cT}) implies that the function
$\ttT (\br, \bs, z)$ is the conformal time of the null ray
$\nu_\br$.  Fixing $\bs$ and $z$, Fermat's principle yields
that the light rays from source to observer are determined by
those null rays in $\cN$ with comoving impact vectors  $\br$
satisfying

\beq
\label{eq-grad-cT}
 \grad_\br \cT (\br, \bs, z)  = \bmo,
\eeq

\noindent where $ \grad_\br$ is the gradient operator in
rectangular coordinates $\br$. By (\ref{eq-cT}) and
(\ref{eq-cT-1}), equation (\ref{eq-grad-cT}) is equivalent to

\beq
\label{eq-lenseq}
\bs
= \frac{z}{z_l}\br
     -(z-z_l)\hat{\bma} (\br) \equiv \bs (\br,z),
\eeq where $\hat{\bma} (\br) \equiv \grad_\br \hp (\br)$ is the
change in direction through which the light signal $\nu_\br$ bends
at the lens plane, and its magnitude is referred to as the 
({\it comoving})  {\it bending angle.}   Equation (\ref{eq-lenseq}) is called the
({\it comoving}) {\it lens equation}. It determines the comoving impact
vectors $\br$ of the light rays from $\bs$ to the observer.

For  fixed $z$,
equation (\ref{eq-lenseq}) defines a map

$$\bs_z: L \rightarrow S: \bs_z (\br) = \bs(\br,z),$$

\noindent called a ({\it comoving})   {\it standard lensing map}  from the lens
plane $L$ to the {\it light source plane} $S$ (i.e., the
$xy$-plane at position $z$ on the optical axis), and carries no
sense of time. This is the comoving version of the cosmological
lensing map commonly used in gravitational lensing --- see
(\ref{eq-bme}).  Note that the Jacobian matrix of $\bs_z$ is symmetric:
\beq
\label{eq-jac-sz}
(\jac \bs_z ) (\br)
     = \frac{z}{z_l} {\bf I}
     -(z-z_l) [\Hess_\br \hp] (\br).
\eeq
Here ${\bf I}$ is the 2$\times$2 identity matrix
and $\Hess_\br \hp$
is the Hessian matrix of
of $\hat{\Psi}$ relative to $\br = (r_1, r_2)$.

The present proper length of each light ray in $\cN$
depends on the comoving position $\bs$ of the source on the source
plane. 
Inserting (\ref{eq-lenseq}) in (\ref{eq-cT-1}),  we
obtain an expression for the present proper length:

\beq
\label{eq-trav-time-light}
\cT (\br, z)\equiv
\frac{a_0}{c} \ttT (\br, z),
\eeq
where

\beqa
\label{eq-trav-time-light-c}
\ttT (\br, z)
& \equiv &
\ttT (\br, \bs (\br,z), z)\nonumber
\\
& =&
z+\frac{1}{2}
    \Big[
      z\frac{|\br|^2}{z_l^2}
     +(z-z_l)
     \Big(|\hat{\bma} (\br)|^2 \nonumber \\
& &  \hspace{0.6in}  - \ 2\frac{\br \cdot\hat{\bma}(\br)}{z_l}
    \Big) \Big]
 \ - \hat{\Psi}(\br)
\eeqa

\noindent with $\bs (\br, z)$ given by (\ref{eq-lenseq}). {\it The
function $\cT(\br, z)$ has the convenient property that the source
coordinate $z$ appears linearly.} We will take advantage of this
fact later in this work.

Notice that though equations (\ref{eq-cT-1}), (\ref{eq-lenseq}),
and (\ref{eq-trav-time-light}) are stated for $z> z_l$,  they
actually hold for all $z$ if  we assume that $\hat{\Psi} (\br)
\equiv 0$ when $-\infty <  z \le z_l$. For example, if $z =0$,
then $\cT (\br, 0) = 0$. If a light ray starts out from a source
at position $(\bs, z)$ in $E$ with $z_l \le z < \infty$, then
$\br$ is the light ray's comoving impact vector  on the lens plane
$L$ at position $z_l$. For $ 0 <  z <   z_l$,  the vector $\br$ is
the comoving position  on $L$ of the light ray's spatial path when
extended backward from $(\bs, z)$ to $L$. If $ - \infty < z < 0$,
then $\br$ is the ray's impact position on $L$ when the light ray
is extended forward from $(\bs, z)$ to the observer to $L$. Since
the case that is of interest to us is when the wavefront has
passed through the lens,  {\it we shall assume, unless stated to
the contrary, that $z\ge z_l$.}

\subsubsection{Present Proper Length Function via Angular Diameter Distances,
Redshifts, and Proper Vectors}

In the majority of the gravitational lensing literature, the
comoving coordinates of the previous sections have not been
exploited. In order to make a connection with standard
gravitational lensing, we now show that the present proper length function
$\cT (\br,\bs,z)$ in (\ref{eq-cT}), which uses comoving
coordinates, can be expressed in terms of the FL angular diameter
distances, redshifts, and proper impact vectors commonly used in
gravitational lensing.

Equations (\ref{eq-cT-gen}) and (\ref{eq-L-ells}) yield

\beq
\label{eq-cT-2}
\cT (\br, \bs, z) = \frac{a_0}{c} \left[(\ell_l + \ell_{ls} - \ell_s)
     - \hat{\Psi} (\br) \right] \ + \ \frac{a_0 \ell_s}{c} .
\eeq

\noindent Since  $a_0 \equiv a(\ttt (\tau_0)) =  a (\tau_0) =
(1+ \ttz_S) a (\tau_S)$ and the FL angular diameter distance
from observer to source is $d_S = a (\tau_S) \ell_s$ for $\kk
=0$, we get

\beq
\label{eq-a-ell}
\frac{a_0 \ell_s}{c} = \frac{1 + \ttz_S}{c} d_S.
\eeq

\noindent The first term on the right of (\ref{eq-cT-2}) is the
time-delay of $\nu_\br$ relative to the FL light ray from the
source to observer in the absence of the lens.  Using the
formula for the time delay  (e.g., p. 74 of \cite{PLW} and
p. 146, \cite{SEF}), we obtain

\beqa
\label{eq-cT-3}
\cT (\br, \bs, z)
& = & \frac{1 + \ttz_L}{c} \frac{d_L d_S}{d_{L,S}}
\Big[
\frac{1}{2} \Big|  \frac{\sfr}{d_L} - \frac{\sfs}{d_S} \Big|^2
\nonumber\\
& & \hspace{.7in}      \ - \ \frac{d_{L,S}}{d_L d_S} \Psi (\sfr)
\Big] \ + \  \frac{1 + \ttz_S}{c} d_S \nonumber \\
& \equiv&   \ \cT (\sfr,
\sfs, d_S), 
\eeqa 
where $\ttz_L$ is the redshift of the lens; $d_L$
is the  FL angular diameter distance from observer to lens and
$d_{L,S}$ the angular diameter distance from lens to source; $\sfr
= a(\tau_L) \br$ is the proper impact vector of the ray at the
lens plane at approximately cosmic time $\tau_L$ and $\sfs =
a(\tau_S) \bs$ is the proper vector of the  source relative to the
optical axis at cosmic time $\tau_S$; and $\Psi_L (\sfr) \equiv
a_L \hat{\Psi} (\br)$ with $a_L = a (\ttt (\tau_L))$.  Note that
$\Psi_L$ has units of length and is the two-dimensional
cosmological potential at cosmic time $\tau = \tau_L$.  Fixing
${\sfs}$ and $d_S$, Fermat's principle yields that light rays are
determined by   $\grad_{\sfr} \cT (\sfr, \sfs, d_S) = \bmo.$
Equation (\ref{eq-cT-3}) now shows that the latter is equivalent
to the usual (cosmological) lens equation:

\beq
\label{eq-sfs}
\sfs = \frac{d_S}{d_L} \sfr \ - \ d_{L,S}\ \grad_{\sfr} \Psi (\sfr).
\eeq

\noindent This induces the standard {\it cosmological lensing map}
in the gravitational lensing literature (e.g., \cite{PLW}, p.77):

\beq
\label{eq-bme}
\bme:\sfL \rightarrow \sfS : \ \bme (\sfr) =
\frac{d_S}{d_L} \sfr \ - \ d_{L,S}\ \grad_{\sfr} \Psi (\sfr),
\eeq

\noindent where   $\sfL = \{a(\tau_L) \br: \br \in L\}$  and $\sfS
= \{a(\tau_S) \bs: \br \in S \}$ are the cosmological lens plane
and light source planes, respectively.

\section{Lightcones, Wavefronts, and Wavefront Lensing Maps}

Consider a gravitational lensing situation with an observer
receiving light from sources beyond the lens plane where
deflectors are located. Even though each light source emits
perhaps in all directions, generating their own lightcones, the
observer receives only those rays that belong to his past
lightcone. Consequently, we shall study the observer's past
lightcone, rather than a source's future lightcone. We are also
interested in constant-time sections of the observer's past
lightcone, which we refer to as wavefronts. In this section, we
shall determine parametric equations  for the past lightcone and
for the wavefronts generating the lightcone.

\subsection{Lensed Lightcones}

The observer's past lightcone $\cL^-$ in  the spacetime $(I\times
E, \bg_{GL})$ is the subset of $I\times E$ consisting of all
past-pointing light rays originating from the observer at
$(\tau_0, 0,0,0)$. Equivalently, the lightcone $\cL^-$ is the set
of all spacetime events in  $(I\times E, \bg_{GL})$ from which a
light ray arrives at the observer at cosmic time $\tau_0$.   To
obtain a formula for the points in $\cL^-$, consider a light ray
$\nu_\br$ starting out at event $(\tau_S, \bs_s, z_s)$, where
$z_s$ may be positive or negative, and arriving at the observer at
$(\tau_0, 0,0,0)$.  Here $\br$ is the  light ray's comoving impact
position on the lens plane $L$ at $z_l$, resulting possibly from
extending the ray backwards or forwards to $L$ ---  see discussion
at end of Section~\ref{subsub-T-le}. Then the (comoving) lens
equations (\ref{eq-lenseq}) and present proper length equation
(\ref{eq-trav-time-light}), show that $\nu_\br$ can be expressed
as follows with $z$ as parameter:

\beq
\label{eq-para-light}
\nu_\br (z) = (\tau_0 - \cT (\br, z), \bs (\br,z), z),
\eeq

\noindent where $z$ varies from $z_s$ to $0$, $\cT (\br,0) =0$,
and $\bs (\br,z_s) = \bs_s$. Alternatively, equation
(\ref{eq-para-light}) holds for a light ray from the observer
to $(\tau_S, \bs_s, z_s)$ by varying $z$ from $0$ to $z_s$.
Hence, by allowing $z$ and $\br$ to vary, we obtain the
following expression for the observer's past lightcone:

\beq
\label{eq-past-lightcone}
\cL^- = \{ (\tau_0 - \cT (\br, z),\ \bs (\br,z),\ z): \ z \in \RR, \
          \br \in L \}.
\eeq

\noindent The set $\cL^-$ is a three-dimensional hypersurface  in
the spacetime  $(I\times S, \bg_{GL})$ and typically has
singularities due to distortions caused by the lens.  In
Section~\ref{sec-caustics-wf-lc}, we shall give equations that
characterize these singularities.   Notice that  in the
conformally equivalent spacetime $(I\times S, \sfg_{\tt GL})$,
where

$$
\sfg_{\tt GL} \equiv \frac{\bg_{\rm GL}}{a^2(\ttt)}
=
-\left (1+2\frac{\phi}{c^2}\right )
                 d \ttt^2 +
        \left (1-2\frac{\phi}{c^2}\right )d S^2_0,
$$

\noindent the observer's past lightcone becomes

\beq
\label{eq-c-L}
\ttL^- = \{ (\ttT_0 - \ttT (\br, z),\ \bs (\br,z),\ z): \ z \in \RR, \
          \br \in L \},
\eeq

\noindent where $\ttT_0 = c \tau_0/a_0.$
The observer's future lightcone $\cL^+$  and conformal
future lightcone $\ttL^+$ are given respectively as follows:
\beqan
\cL^+ &=& \{ (\tau_0 + \cT (\br, z),\ \bs (\br,z),\ z): \ z \in \RR, \
          \br \in L \},\\
\ttL^+ &=&  \{ (\ttT_0 + \ttT (\br, z),\ \bs (\br,z),\ z): \ z \in \RR, \
          \br \in L \}.
\eeqan

\subsection{Lensed Wavefronts and Wavefront Lensing Maps}

With fixed $z$, we saw that equation (\ref{eq-lenseq}) defines
a lensing map from the lens plane to the light source
plane at $z$. There is an alternative interpretation of the
lensing map stemming from the lightcone point of
view.  We can consider (\ref{eq-lenseq}) and
(\ref{eq-trav-time-light}) jointly with the condition of fixed
conformal travel time, instead of fixed $z$.

We define a {\it comoving} ({\it optical})  {\it wavefront} in
$E$ as a surface of constant conformal present proper length $\ttT (\br, z)$. One should keep
in mind that conformal time flows at a rate different from that
of cosmological time --- see equation (\ref{eq-timerates}).
However, the surfaces of constant conformal  time $\ttT$
are the same as those of constant present proper length 
$\cT$; they are only labeled differently --- see
(\ref{eq-trav-time-light}).  Since we are not particularly
interested in the specific labels of the wavefronts, which are
not observable, we prefer to use the conformal time
$$ T \equiv \ttT (\br, z)$$
to describe the wavefronts.
We shall refer to $\ttT$ as a conformal present proper length function.

Due to the linearity of (\ref{eq-trav-time-light-c})  in the
variable $z$, we obtain a parametric expression in terms of
$(\br, T)$ for all the $z$-coordinates of the  comoving
wavefront by solving for $z$:

\beq
\label{eq-zwavefront}
     z(\br, T)
    = \frac{\displaystyle T
        \  - \  \br \cdot \hat{\bma} (\br)
        \  +  \  \frac{z_l}{2} |\hat{\bma} (\br)|^2
                 \  + \  \hat{\Psi} (\br)
                }{\displaystyle
         1
        \  + \  \frac{|\br|^2}{2 z_l^2}
        \  +  \  \frac{|\hat{\bma} (\br)|^2}{2}
        \  - \  \frac{\br \cdot \hat{\bma} (\br)}{z_l}
            }.
\eeq
The values of $z (\br, T)$ are assumed to obey $z
(\br, T) \ge z_l$.  In other words,  we are interested
primarily in the lensed portion of the front, i.e., the portion
of the front with $z$-values on the optical axis from the lens
plane onward.   Note that if we allow  $z (\br, T) < z_l$, then
it is assumed that $\hat{\Psi} (\br) \equiv 0$ in
(\ref{eq-zwavefront}) --- see the end of
Section~\ref{subsub-T-le}.

Equation (\ref{eq-zwavefront}) can now be used in
(\ref{eq-lenseq}) to obtain an expression for the points of the
wavefront that are on the $xy$-plane at position $z (\br, T)$:
\beqa
\label{eq-swavefront}
\bs (\br, T)
& \equiv &
\bs (\br, z (\br, T)) \nonumber \\
& =& 
z (\br, T)  \left(\frac{\br}{z_l}
     \  - \   \hat{\bma}(\br)\right)
\  +  \  z_l  \hat{\bma} (\br).
\eeqa
In other words, the pair $(\bs, z)$ determined by
(\ref{eq-swavefront}) and (\ref{eq-zwavefront}) defines a chart
(possibly with singularities) on the portion of the wavefront
that went through the lens plane.  Of course, the pair can
also  be used as a chart for the portion of the front behind
the lens plane (e.g., by setting  $\hat{\Psi} (\br) \equiv 0$). The
{\it comoving wavefront} at conformal time $T$ is then given by

\beq
\label{eq-wf}
\cW (T) \equiv \{ (\bs(\br,T), z(\br, T)): \br \in L\} \subseteq E,
\eeq

\noindent which is a  two-dimensional surface that generically
has singularities due to lensing.   Notice that if we allow  $
- \infty < z < \infty$ and $0\le T \le T_0$, then $\cW (T)$
traces out the conformal light cone $\ttL^-$ in
(\ref{eq-c-L}).

For the cosmological present proper length 

$$t \equiv \cT (\br, z),$$

\noindent the wavefront is given  in the spacetime $(I \times
E, \bg_{GL})$ at cosmic time $\tau_0$ by

$$
\cW_0 (t) \equiv \{ (\tau_0 -  t, \bs(\br,t),  z(\br, t)): \br \in L\} \subseteq I \times E,
$$

\noindent where $\bs(\br,t) \equiv \bs(\br, T (t))$ and $z(\br,
t) \equiv z (\br, T (t))$ with $T(t) = c t/a_0.$ Equation
(\ref{eq-past-lightcone}) shows that  $\cW_0 (t)$ is a subset
of the past lightcone $\cL^-$. The wavefront $\cW_0 (t)$ is the
set of positions of all sources  whose light rays left at the
same cosmic time $\tau_0 - t$ and arrive at the observer at
roughly the present cosmic time $\tau_0$.  Most observed lensed
light sources are at distances ranging from of order tens of
kiloparsecs (e.g., microlensing) to Gigaparsecs (e.g., quasar
lensing).  The  observation time is typically of order weeks
to  a few years.   Consequently, we expect the wavefront $\cW_0
(t)$ to change negligibly during the observing time, unless the
front happens to pass through a higher order singularity during
the observation.

The wavefront $\cW_0 (t)$ lies in the proper space $\{\tau_0 -
t\} \times E$.  Projecting the front into the comoving space,
we obtain

$$
\cW^E_0 (t) \equiv \{ (\bs(\br,t),  z(\br, t)): \br \in L\} \subseteq E,
$$

\noindent which is a constant cosmic time slice of $\cL^-$
(namely, $\tau_0 - t = {\rm constant}$). The wavefront $\cW^E_0
(t)$ differs from  $\cW (T)$ in $E$ merely by the label $T = c
t/a_0.$ The two wavefronts are isometric as singular spaces in
the comoving space with the Euclidean metric.

Now for fixed $T$, the lens equation (\ref{eq-lenseq})
determines a mapping $$\bw_T: L \rightarrow \cW (T) \subseteq
\RR^3,$$ called a {\it comoving wavefront lensing map}, defined
by

$$\bw_T (\br) = (\bs(\br,T),  z(\br, T)).$$

\noindent By allowing for $z$-values with $z(\br, T) < z_l$,
the map $\bw_T$ can be separated into two single-valued maps

$$ \bw_T (\br) = \left\{
                  \begin{array}{ll}
                     \bw^+_T (\br) & \quad  {\rm if} \ z(\br, T) \ge z_l,  \\
                      \bw^-_T (\br) & \quad  {\rm if} \ z(\br, T) < z_l,
                  \end{array}
                 \right.
$$

\noindent where $\bw^+_T: L \rightarrow \cW^+ (T)$ and
$\bw^-_T: L \rightarrow \cW^- (T)$ with $ \cW^+ (T)$  the
subset of $\cW (T)$ for which $z \ge z_l$,  while $\cW^- (T)$
is the subset with $z < z_l$.   In other words, the map $\bw_T$
defines two (possibly singular) coordinate patches on $\cW
(T)$.  Since  $\cW^- (T)$ did not pass through the lens, it has
no singularities and so the map $\bw^-_T$ is of little interest
for our purposes. Instead,  we shall focus primarily on
$\bw_T$.  {\it Unless stated to the contrary, we shall assume
that $\bw_T (\br) = \bw_T^+ (\br).$}

It is important to emphasize that rather than the usual mapping
from a lens plane to the light source plane (such as given by the
comoving and cosmological lensing maps $\bs_z: L \rightarrow S$
and $\bme: \sfL \rightarrow \sfS$), {\it we now have a new mapping
--- one from the lens plane to a wavefront.} This gives us an
alternative interpretation of gravitational lensing where the
light source plane is dispensed with and substituted with a
wavefront: the locus of points that can be reached in a given time
by light signals emitted simultaneously from the observer in all
possible directions, after passing through the lens plane. It
should also be kept in mind that even though the wavefront as a
surface is different from a source plane, in typical lensing
scenarios the wavefront surface lies very close to a plane
in the region of interest, namely around the optical axis.

\section{Caustics on Wavefronts and Lightcones}
\label{sec-caustics-wf-lc}

As a surface in the comoving space $E \subseteq \RR^3$, the
wavefront $\cW^+ (T)$ typically develops singularities ---
called {\it caustics} --- due to the distortions caused by the
lens.  More precisely,  a point $\br$ in the lens plane $L$ is
a {\it critical point} of the wavefront lensing map

$$\bw_T: L \rightarrow \cW^+ (T) \subseteq \RR^3, \qquad
\bw_T (\br) = (\bs(\br,T), z(\br, T))$$

\noindent if $\rank [ (\jac \bw_T) (\br)] < 2,$ where $\jac
\bw_T$ is the Jacobian matrix of $\bw_T$ and $z(\br, T) \ge
z_l$. The set of critical points of $\bw_T$ will be denoted by
$\crit (\bw_T)$. The set of {\it caustic points} of $\bw_T$ or
on $\cW^+ (T)$ is the set $$\caustics [\cW^+ (T)] \equiv \bw_T
[ \crit (\bw_T)]$$ of all critical values of $\bw_T$.

The Jacobian matrix of $\bw_T$ is the following $3\times 2$
matrix:

\beq
\label{eq-jac-matrix}
\jac \bw_T =  \left[ \begin{array}{l}
        {\displaystyle \frac{\pa \bs (\br, T)}
             {\partial \br} }   \\
                            \\
        {\displaystyle    \frac{\pa z (\br, T)}
             {\pa \br }}
        \end{array}
    \right],
\eeq

\noindent where  $z = z (\br, T)$ is given by
(\ref{eq-zwavefront}) and  $\bs = \bs (\br, T)$ by
(\ref{eq-swavefront}).
Define a  function $f: L \rightarrow \RR$
by
\beq
\label{eq-f}
f (\br) = z_l \ + \ \frac{|\br|^2}{2 z_l} - \hat{\Psi} (\br),
\eeq
Note  that $f (\br)  = \ell_l - \hat{\Psi} (\br)$
within our approximations (see (\ref{eq-ells})),
so $a_0 f (\br)/c$ is the present proper length of a light ray
from the point
$\br$ on the lens plane to the observer.
By (\ref{eq-lenseq}), we see that
$$
\bs (\br, T) =
\bs (\br, z (\br, T))=
                z (\br, T) \ \nabla_\br f (\br) \ + \  z_l \hat{\bma} (\br)
$$
Consequently, the Jacobian matrix 
of
$\bs (\br, T)$ relative to $\br$ can then be expressed
as
\beq
\label{eq-dsdr}
{
\frac{\pa \bs}
                     {\partial \br} }
 =  \left[ \begin{array}{lr}
{
f_1 \ \frac{\pa z}{\pa r_1}  + z f_{11} + z_l  \hat{\Psi}_{11}  }
&
\
{
f_1 \ \frac{\pa z}{\pa r_1}  + z f_{12} + z_l \hat{\Psi}_{12} }
\\
& \\
{
f_2 \ \frac{\pa z}{\pa r_1}   + z f_{21} + z_l \hat{\Psi}_{21}  }
& 
\
{
f_2 \ \frac{\pa z}{\pa r_2}   + z f_{22} + z_l \hat{\Psi}_{22} }
\end{array}
        \right], 
\eeq
where
$f_i$, $f_{ij}$, and
$\hp_{ij}$ denote the usual
partial derivatives
relative to $\br = (r_1, r_2).$
Note that the Jacobian  matrix in (\ref{eq-dsdr}) is {\it not} symmetric,
unlike  the usual
Jacobian matrices $\pa \bs_z/\pa \br$ and $\pa \bme/\pa \sfr$
of the comoving and cosmological lensing maps. If $z$ is fixed,
then (\ref{eq-dsdr}) reduces to the usual comoving symmetric
case --- see equation (\ref{eq-jac-sz}):
$$
\frac{\pa \bs (\br, T)}{\partial \br} = \frac{\pa \bs_z (\br)}{\partial \br}
=  \frac{z}{z_l} {\bf I}
     - ( z - z_l)\ (\Hess_\br \hp) (\br).
$$

We have  $\rank [ (\jac \bw_T) (\br)] < 2$ if and only if every
$2$-square minor of $\jac \bw_T$ vanishes:

\beqa
\label{eq-2-minor-a}
\det {\displaystyle \frac{\pa \bs}{\pa \br}} & = & 0,\\
\nonumber \\
\label{eq-2-minor-b}
\det \left[ \begin{array}{l}
                {\displaystyle \frac{\pa s_1}
                     {\pa \br} }   \\ \\
                {\displaystyle    \frac{\pa z}
                     {\pa \br }}
                \end{array}
        \right]
      & = & \frac{\pa s_1}{\pa r_1} \ \frac{\pa z}{\pa r_2}
             \ - \ \frac{\pa s_1}{\pa r_2} \ \frac{\pa z}{\pa r_1} = 0,\\
\label{eq-2-minor-c}
\det \left[ \begin{array}{l}
                {\displaystyle \frac{\pa s_2}
                     {\pa \br} }   \\ \\
                {\displaystyle    \frac{\pa z}
                     {\pa \br }}
                \end{array}
        \right]
      & = & \frac{\pa s_2}{\pa r_1} \ \frac{\pa z}{\pa r_2}
             \ - \ \frac{\pa s_2}{\pa r_2} \ \frac{\pa z}{\pa r_1} = 0,
\eeqa
where $\bs = (s_1,s_2).$   Since
the conformal time $T$  is fixed when  studying $\cW^+ (T)$, we
have

$$\ttT (\br, \bs (\br, T), z (\br, T)) = T = \ {\rm constant},$$

\noindent which yields

\beq
\label{eq-dzdr-a}
\frac{\pa \ttT}{\pa \br} \ + \ \frac{\pa \ttT}{\pa \bs} \ \frac{\pa \bs}{\pa \br}
 \ + \  \frac{\pa \ttT}{\pa z} \ \frac{\pa z}{\pa \br}
= \bmo.
\eeq

\noindent By (\ref{eq-grad-cT}), we get

\beq
\label{eq-grad-ttT}
\frac{\pa \ttT}{\pa \br}  = \bmo,
\eeq

\noindent while equation (\ref{eq-cT-1}) yields $\pa \ttT/\pa
\bs = (\bs - \br)/(z - z_l)$. The latter  vanishes if and only
if  $\hat{\bma} (\br) = \br/z_l$ (apply (\ref{eq-grad-ttT}) or
(\ref{eq-lenseq})), which holds if and only if $\br$ is a
critical point of the function $f$ in
(\ref{eq-f}).
Note that for $\br$ to be a critical point of $\xi$ it must
solve

$$\frac{\br}{z_l} \ - \ \hat{\bma} (\br) = \frac{\pa f}{\pa \br} (\br) = \bmo.
$$

\noindent Generically, the critical points of $f$ are
nondegenerate and so are isolated points
(e.g., p. 240, Petters et al.).  For this reason, {\it
we shall assume ---  unless stated to the contrary --- that
$\br \neq \bs$.}  Consequently,

\beq
\label{eq-dTds-not-zero}
\frac{\pa \ttT}{\pa \bs}  = \frac{\bs - \br}{z - z_l}\neq \bmo,
\qquad
\frac{\pa \ttT}{\pa z}  = -\frac{|\br - \bs|^2}{2(z - z_l)^2}\neq 0.
\eeq

\noindent Equation (\ref{eq-dzdr-a}) yields

\beq
\label{eq-dzdr-b}
\frac{\pa z}{\pa \br}
= \left[\frac{(\pa \ttT/\pa \bs)}{(\pa \ttT/\pa z)} \right]
  \ \frac{\pa \bs}{\pa \br},
\eeq

\noindent where by (\ref{eq-dTds-not-zero}) the bracketed term
is nonzero. Plugging $\pa z/\pa r_1$ and $\pa z/\pa r_2$ from
(\ref{eq-dzdr-b}) into (\ref{eq-2-minor-b}) and
(\ref{eq-2-minor-c}) implies that

\beqa
\label{eq-2-minor-b-2}
\hspace{-0.5in}
\frac{\pa s_1}{\pa r_1}\frac{\pa z}{\pa r_2}
            - \frac{\pa s_1}{\pa r_2} \frac{\pa z}{\pa r_1}
& = &  
\left[\frac{(\pa \ttT/\pa s_2)}{(\pa \ttT/\pa z)} \right]
  \det \frac{\pa \bs}{\pa \br},\\
\label{eq-2-minor-c-2}
\hspace{-0.5in}
\frac{\pa s_2}{\pa r_1} \frac{\pa z}{\pa r_2}
             - \frac{\pa s_2}{\pa r_2} \frac{\pa z}{\pa r_1}
& = & 
- \left[\frac{(\pa \ttT/\pa s_1)}{(\pa \ttT/\pa z)} \right]
   \det \frac{\pa \bs}{\pa \br}.
\eeqa

\noindent Equations (\ref{eq-2-minor-a}) --
(\ref{eq-2-minor-c}), (\ref{eq-2-minor-b-2}),
(\ref{eq-2-minor-c-2}), along with our assumption  in
(\ref{eq-dTds-not-zero}), then show that the Jacobian matrix of
$\bw_T$ has rank below $2$ if and only if the Jacobian
determinant of $\bs$ vanishes. Thus, at the conformal time
$T$ the caustics on the
wavefront are given by
\beq
\label{eq-caustics}
\caustics [\cW (T)] 
=\Big\{ (\bs (\br, T), z(\br, T ))\Big\}, 
\eeq
where $\br \in L$, $\bs (\br, T)$
is given by (\ref{eq-swavefront}) with
$T$ fixed, and $\bs = \bs (\br, T)$ is subjected to the constraint
\beq
\label{eq-c-constraint}
\det
\frac{\pa \bs}{\pa \br}
                          (\br, T) = 0.
\eeq
Equation
(\ref{eq-caustics}) extends to wavefront lensing maps the usual
concept of  caustics in gravitational lensing, where  lensing
maps are from a lens plane to a light source plane. Allowing
the conformal time $T$ to vary in (\ref{eq-caustics}) yields
evolving caustics on the wavefront as the front propagates. The
caustics on a wavefront trace out a {\it caustic surface} on
the conformal past lightcone $\ttL^-$ is given by

\beq
\label{eq-conf-caustic-surface}
\caustics [\ttL^-] 
= \Big\{ (T_0 - T (\br), \bs (\br, T (\br)), z(\br, T (\br) ))\Big\}
\eeq

\noindent
where $T_0 = c
\tau_0/a_0$, $\br \in L$, and $\bs (\br, T)$ obeys
(\ref{eq-c-constraint}).
Since $T$ varies in the case of a caustic surface, 
the vanishing-determinant condition 
(\ref{eq-c-constraint}) forces $T$ to be a function
of $\br$, which we have expressed by $T(\br)$.
Also, note that the caustics lie on the portion of
$\ttL^-$ with $z \ge z_l$.  Projecting $\caustics [\ttL^-]$
into the comoving space $E$, we obtain the {\it comoving
caustic surface}
\beq
\label{eq-comov-caustic-surface}
\caustics_E [\ttL^-]
= \Big\{ (\bs (\br, T (\br)), z(\br, T (\br) )) \Big\},
\eeq
where $\br \in L$ and $\bs (\br, T (\br))$ satisfies (\ref{eq-c-constraint}).
In the
spacetime $(I \times E, \bg_{GL})$, the caustic  is given by

\beq
\label{eq-caustic-surface}
\caustics[\cL^-]
 = \Big\{ (\tau_0 - \tau (\br), \bs (\br, T(\br)), 
z(\br, T(\br) ))\Big\}
\eeq
where $\tau_0$ is the present cosmic time and
$\tau (\br) \equiv
a_0 T (\br)/c$, $\br \in L$, and
$\bs (\br, T(\br))$ obeys (\ref{eq-c-constraint}).
 Projecting $\caustics[\cL^-]$ into $E$ yields a
surface that is  isometric to $\caustics_E [\ttL^-]$ as singular spaces
relative
to the Euclidean metric on $E$.  Analogous to equations
(\ref{eq-conf-caustic-surface}) --- (\ref{eq-caustic-surface}),
the caustics of the
future lightcone and their projection into the comoving space are given as follows:
\beqa
\label{eq-conf-caustic-surface-f}
\caustics[\cL^+]  & = & \Big\{ (\tau_0 + \tau (\br), \bs (\br, T(\br)), z(\br, T(\br) ))
\Big\},
\nonumber \\
&& \\
&&
\nonumber \\
\label{eq-caustic-surface-f}
\caustics [\ttL^+] & = & \Big\{ (T_0 + T (\br), \bs (\br, T (\br)), z(\br, T(\br) ))
\Big\},
\nonumber \\
&& \\
&&
\nonumber \\
\label{eq-comov-caustic-surface-f}
\caustics_E [\ttL^+] & = & \Big\{ (\bs (\br, T (\br)), z(\br, T (\br) ))
\Big\}, 
\eeqa
where $\br \in L$ and 
$\bs (\br, T )$ satisfies (\ref{eq-c-constraint}).

The previous discussion traces out the caustic surfaces of an
observer's past lightcone using constant time slices.
The caustic surfaces can also be traced out using constant
$z$-slices.
Slices of the caustic sheet by constant $z$-planes are curves
on the light source plane, and coincide with what are commonly referred to in the lensing literature
as ``caustics.''
We shall refer to such caustics
as {\it $z$-sliced caustic curves.}
These caustic
curves can also have singularities, such as cusps.
It is important to add that our discussion shows that points on
a $z$-sliced caustic curve actually occur at different cosmic (or conformal) times.
Often times, points on caustic curves are treated in the lensing
literature as if they occur at the same cosmic time.

In order to calculate the points on the comoving caustic sheet,
we impose the condition that the
Jacobian of the lens map be singular --- see
(\ref{eq-comov-caustic-surface}) and
(\ref{eq-comov-caustic-surface-f}):
$$\det \frac{\pa \bs}{\pa \br} =  0.$$
This condition
can be imposed either at constant $z$ or constant $T$ if the
intention is to produce the caustic sheet.
For constant $T$, we have that $\bs$ is given by
the wavefront lensing map (\ref{eq-swavefront}),  i.e., $\bs = \bs (\br, z(\br,T))$,
while for constant $z$
the function $\bs$ is given by the standard lensing map (\ref{eq-lenseq}),
i.e., $\bs = \bs_z (\br)$.
In the latter case,
we consider the caustics of the mapping $\bs = \bs_z$
from the lens plane to the light source
plane, with the intention of eventually letting the light source
plane at $z$ sweep (along the optical axis) the entire range behind the lens.

By equation (\ref{eq-jac-sz}) ---  equivalently, equation
(\ref{eq-dsdr}) with $z$ fixed  ---
we see that the Jacobian matrix of $\bs_z$
depends linearly on $z$:
\beq
\label{eq-dszdr}
{\displaystyle \frac{\pa \bs_z}
                     {\partial \br} }
=
\left[ \begin{array}{lr}
{\displaystyle   1 + (z - z_l)  f_{11} }
& \qquad
{\displaystyle  - (z - z_l) \hp_{12} }
\\  \\
& \\
{\displaystyle  - (z - z_l) \hp_{21} }
& \qquad
{\displaystyle  1 +  (z - z_l)  f_{22} }
\end{array}
        \right],
\eeq
where $f$ is given by (\ref{eq-f}).
Consequently, the
determinant of (\ref{eq-dszdr}) is a quadratic in $z$. Explicitly,
the vanishing of the Jacobian determinant of $\bs_z$ is
\beqa
\label{eq-det-dszdr}
    0
  & =&  1
 +   (z -  z_l)
     (2z_l^{-1} - \hat{\Psi}_{11}-\hat{\Psi}_{22})
      +    (z -  z_l)^2
     \Big[(z_l^{-1} 
\nonumber \\
&& \hspace{1.2in}
-   \hat{\Psi}_{11})
           (z_l^{-1} -   \hat{\Psi}_{22})
                     -   \hat{\Psi}_{12}^2\Big].
\nonumber \\
&&
\eeqa
There are two solutions for $z$ as a
function of $\br$, representing the $z-$coordinate of points on
the caustic surface:
\beqa
\label{eq-zpm}
\hspace{-0.2in}\mbox{}
&& z_\pm (\br)
  \equiv z_l
      + 
  \frac{\hat{\Psi}_{11} +  \hat{\Psi}_{22}  -   2z_l^{-1}
           \pm 
        \sqrt{(\hat{\Psi}_{11}  -    \hat{\Psi}_{22})^2
          +    4\hat{\Psi}_{12}^2}}
       {2[(z_l^{-1}  -   \hat{\Psi}_{11})
      (z_l^{-1}  -   \hat{\Psi}_{22})
                -   \hat{\Psi}_{12}^2]}.
\nonumber \\
\hspace{-0.4in}\mbox{}&&
\eeqa
Evaluating $\bs_z = \bs(\br, z)$ at $z_\pm (\br)$, we obtain
the transverse coordinates of points on the caustic surface:
\beq
\label{eq-spm}
 \bs_{\pm}(\br) \equiv \bs(\br, z_\pm (\br))
= \frac{z_\pm (\br)}{z_l}\ \br \ - \  \Big( z_\pm (\br) \ - \ z_l \Big)
\ \hat{\bma} (\br).
\eeq

\noindent
By varying $\br$ across the lens plane,
the pair $(\bs_\pm (\br), z_{\pm}(\br))$ traces out
the comoving caustic surface for either the future or
past conformal lightcone:
\beqa
\label{eq-comov-caustic-surface-z}
\caustics_E [\ttL^+] 
&=& \Big\{ (\bs_\pm (\br), z_{\pm}(\br)) :
\   \br \in L\Big\} \nonumber \\
&=& \caustics_E [\ttL^-].
\eeqa

\section{Surfing a Caustic Sheet \label{sec-causticsurfing}}

In his Millennium Essay, Blandford \cite{Bl01} speculated
about some possible novel gravitational lensing ways
of probing the cosmos.  One of these dealt with
caustic sheets:
``Our observations need not be  passive.
.. For example, suppose that we launch  an array of
robotic telescopes and use three of them to
measure the velocity of a caustic sheet (from
by a bright source) as it passes Earth;  a fourth
telescope could be made to ``surf'' the wave and
observe the source with considerable magnification
for a long time.''

We now apply the results of the previous section to
show how a telescope may surf a sheet of a caustic surface.
This will be done in two steps.  

First, change the point of
view by assuming that the telescope is the observer.
We shall then consider the past comoving caustic surface of the telescope
and the future comoving caustic surface of the source simultaneously,
and determine the analytical form of the equations for these surfaces.
Assume that the telescope lies somewhere on the future comoving
caustic surface of the source (i.e., the source is seen
at an extremely high magnification corresponding to being
on a caustic).  For the time being, we shall also suppose that the source is
at rest relative to the lens.  The past comoving caustic surface
of the telescope is given by an expression of the form
$$S_0 (\sfx) =0,$$
where $\sfx = (\bs,z)$.   In principle, one can find $S_0$ by solving for $\br$ 
in terms of $\bs_\pm$ using (\ref{eq-spm}) and inserting
those values of $\br$ into (\ref{eq-zpm}) to obtain
$z_\pm$ in terms of $\bs_\pm$, say,
$z_\pm = F_\pm (\bs_\pm).$ 
In this case, the past comoving caustic surface of the telescope
is given by 
\beq
\label{eq-S0}
S_0 (\sfx) \equiv z_\pm - F_\pm (\bs_\pm) = 0.
\eeq
Since we are using the small-angle approximation, we can approximate
the future comoving caustic surface of the source by
(\ref{eq-S0}) if $\bs$ and $z$ are transformed as follows:
$$\bs \rightarrow - \bs \ \frac{\ell_{tl}}{\ell_{sl}},
\qquad z \rightarrow  \ell_{tl} + \ell_{sl},$$
where $\ell_{tl}$ and $\ell_{sl}$ are the telescope-lens and
lens-source Euclidean distances, resp.
Hence, we suppose that there is an expression for the
future comoving caustic surface of the source of the
form
\beq
\label{eq-implicitcaustic}
    S_0(\sfx) = 0.
\eeq

Second, 
suppose that a source moving relative to the lens with 
slowly varying 3-velocity ${\sf V}$ in comoving
coordinates is observed extremely magnified by a telescope at a
given instant of conformal time, say,
$\sft = 0$.   Assuming that the telescope lies 
somewhere on a caustic sheet of the source away from singularities
(e.g., cusp ridges, swallowtails, elliptic umbilics,
hyperbolic umbilics), what should the
subsequent motion of the telescope be in order to track the image
of the source at a peak brightness? 
Suppose that during a time increment $\delta \sft$
the source moves keeping the distance to the lens plane
approximately the same, and assuming its motion to be relatively
slow, the caustic sheet of the source in motion will not differ
from that of the source at rest other than by a general
translation in the direction of motion, i.e.,
$$
S(\sfx, \delta \sft) =S_0(\sfx-{\sf V}\ \delta \sft)=0.
$$
If the telescope lies on the caustic sheet at point $\sfx_0$ at
time $\sft=0$, then
$$
   S(\sfx_0,0) = 0 =S_0(\sfx_0),
$$
For the telescope to see a bright image of the source
during a length of time $\delta \sft$, the telescope needs to move with
velocity ${\sf v}$ to another point on the caustic sheet, i.e.,
to a point $\sfx = \sfx_0+{\sf v}\delta\sft$ so that
$$
    S(\sfx,\delta \sft) = 0 =
    S_0(\sfx_0 + ({\sf v}-{\sf V})\delta \sft).
$$
Taylor expanding to first order in $\delta \sft$,  this is equivalent to
\begin{equation}\label{eq-constraint}
\nabla S_0\cdot ({\sf v}-{\sf V}) = 0,
\end{equation}
where $\nabla S_0$ is evaluated at the original position
of the telescope $\sfx_0$ and $\cdot$ stands for
the Euclidean scalar product of the two vectors. This means that
the telescope's velocity must differ from that of the source at
most by a vector tangent to the caustic sheet. Clearly ${\sf v} =
{\sf V}$ would be one way to stay on the caustic sheet, but it may
require more energy than necessary.  We want the smallest speed needed to
stay on the caustic sheet.  In order words,
we shall minimize  $|{\sf v}|^2$ for a fixed spatial position and
time subjected to the constraint in (\ref{eq-constraint}).
This will be valid for the time increment $\delta \sft$, i.e.,
we are considering only the zeroth iterate of caustic surfing.
The vanishing gradient of the Lagrangian 
${\cal L} = v_1^2 + v_2^2 + v_3^2 + \lambda \ \nabla S_0\cdot ({\sf v}-{\sf V})$
with respect to ${\sf v} = (v_1, v_2, v_3)$ and
the multiplier $\lambda$ yields 
$\lambda = - {\sf V}\cdot \nabla S_0/|\nabla S_0|^2$.  Hence,
we obtain the minimum velocity
\beq\label{eq-vmin}
{\sf v}_{min} = - \lambda \nabla S_0 = \frac{{\sf V}\cdot\nabla S_0}{|\nabla
S_0|^2}\nabla S_0,
\eeq
where $\nabla S_0$ is evaluated at $\sfx_0$.
In other words, the minimum velocity is the projection of ${\sf V}$ to the
normal vector $\nabla S_0$ to the caustic surface at $\sfx_0$.    This
is physically the minimum speed  since $\nabla S_0$ points along the
shortest direction between the caustic sheet at time $\sft =0$ and
the sheet at $\sft = \delta \sft$.
The speed $|{\sf v}_{min}|$ might be considerably
smaller than the source's speed $|{\sf V}|$ depending on the circumstances. The
normal unit vector to the caustic $\nabla S_0/|\nabla S_0|$ can be
calculated either from the expression (\ref{eq-implicitcaustic})
if available or from the parametric version of the caustic map
given by (\ref{eq-zpm}) and (\ref{eq-spm}).

In a physically realistic situation the telescope will spend some
time to reach this velocity, and one will need to adjust the velocity
continuously even if the source is moving at constant speed, due
to the curvature of the caustic sheet.

We shall illustrate the caustic surfing concept with a
singular-elliptical-potential example in
Subsection~\ref{subsec:sing-ep}.

\section{Examples and Illustrations}

In this section we illustrate explicitly the construction of the
wavefronts in some of the most widely used models for the deflection
potential in astrophysics.   In all the cases that we deal with the
expressions for the deflection potentials stated are assumed to hold
only in the vicinity of the optical axis. This assumption is
maintained in the construction of all the figures. 

\subsection{Nonsingular Elliptical Potential Lens}

In this subsection we illustrate the case of an elliptical potential,
which has the following form:

\begin{equation}
    \hp_{ep} (\br) \equiv
        A_0 \
        \sqrt{r_c^2+(1-\epsilon)r_1^2+(1+\epsilon)r_2^2}
\end{equation}
where $A_0$ is a dimensionless constant. This potential is often
used to model elliptical galaxies with $A_0$ proportional to the
velocity dispersion of the lens, whereas the dimensionless
parameters $r_c>0$ and $\epsilon\ge 0$ determine the core radius
and ellipticity of the lens. A non-vanishing core radius $r_c$
guarantees a non-singular surface mass density, which in turn
ensures that no light rays are obstructed in their passage through
the lens. The bending angle is given by

$$\hat{\bma}_{ep} (\br) 
= \nabla_\br \hp_{ep} (\br) =
\frac{A_0 \ \Big( (1+\epsilon)r_1,\ (1-\epsilon)r_2\Big)}
     {\sqrt{r_c^2+(1-\epsilon)r_1^2+(1+\epsilon)r_2^2}}.
$$
For fixed parameter values of $A_0$, $r_c$, and $\epsilon$,
these two explicit expressions for the potential and bending angle
can be used in $z(\br,T)$ and $\bs(\br,T)$ to produce plots of the
constant-time wavefronts embedded in three-dimensional space
beyond the lens plane.  At early times beyond the lens plane, the
wavefront is essentially spherical, but as it progresses away from
the lens plane the central section of the wavefront begins to lag
behind the rest of the wavefront, eventually folding and
developing multiple sheets. The progression of the wavefront in
time shows different regimes, from smooth to various singular
types. In the following, we produce a representative surface of
each regime, effectively classifying the singularities of the
wavefronts of an elliptical lens.  This extends the brief summary
of such wavefronts given in \cite{FP01}.  Note
that formally our treatment is along the future lightcone of the
observer, which can be interpreted as the past lightcone in a time
coordinate running towards the past, because the spacetime is
static.

\subsubsection{Wavefront Singularities}

We arbitrarily fix the values of the parameters at $A_0 = 1.01,
z_l=90, \epsilon = 0.0002$ and $r_c=1.001$. These values are
chosen for purely pedagogical reasons. Notwithstanding, in making
our choice we took care to ensure that the approximation of small
angles was met, in order for the plots to be qualitatively
representative of lensing problems.

Let $T_l$ be the time at which the ray that passes through the
center of the lens (i.e., the origin) reaches the lens plane. With
our choice of parameters, it takes the value $T_l=90$. We find
that for times at least as late as $T=T_l+8900$, the wavefronts
are regular on the other side of the lens plane, but sometime
before $T=T_l+9800$ the first singularity occurs in the form of a
single point on the wavefront. The time scales in this particular
example are irrelevant; in physically accurate lensing situations
the time scale should agree with the distance scale to the lens
plane in order of magnitude. The time of the first singular
wavefront is about $T=T_l+9699.27$, and is calculated in
Section~\ref{sec:regularcausticsheet}.   After the first
singularity, the wavefront successively goes through four more
distinct singularity regimes.

\begin{figure}
\vbox{
\includegraphics[width=3in,angle=-90]{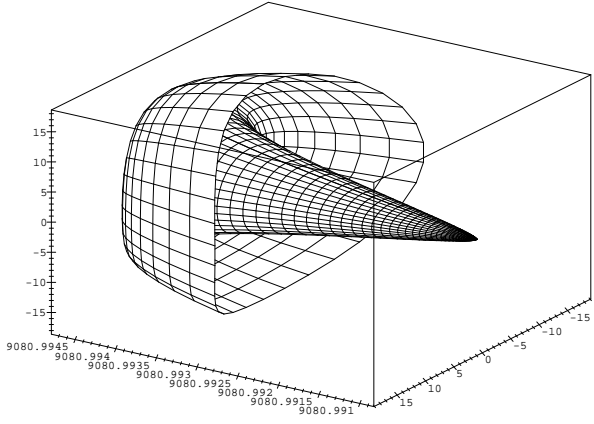}
        }
\caption{\label{fig:spike} A portion of the wavefront of an elliptical potential in the
regular regime, i.e., early times past the lens plane (in this case,
$T=T_l+8900$). The observer is to the right at 
$(0,0,0)$.}
\end{figure}

The regular regime is illustrated in Figure~\ref{fig:spike}, which
shows a field view of the wavefront. The observer is located to
the far right, at (0,0,0) in the plot coordinates. The wavefront
is distorted with respect to a sphere, and develops a dent that
points toward the observer along the optical axis. The scale of
the optical axis is greatly magnified (about $10^4$ times) with
respect to the transversal axes in order to better show the
retardation effect on the wavefront. The ``spike'' has a total
length of about $10^{-3}$ on a sphere of radius about $9\times
10^3$, so the effect is small in the proper scales and is well
within the approximation of small angles.  The tip of the
``spike'' is perfectly smooth and presents a convex surface to the
observer. There are no singularities anywhere in the wavefront at
this time ($T=T_l+8900$).

\begin{figure}
\vbox{
\includegraphics[width=3in,angle=-90]{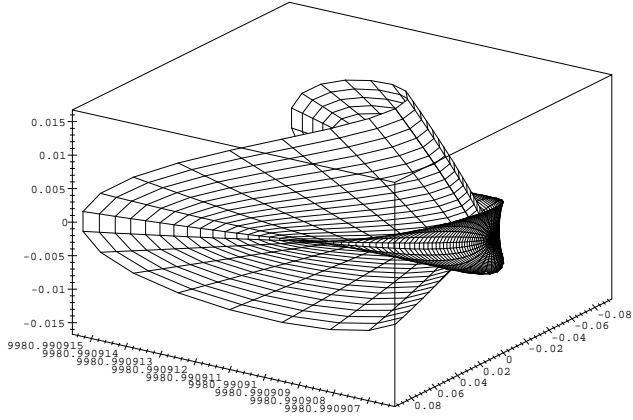}\\
\includegraphics[width=2.5in,angle=-90]{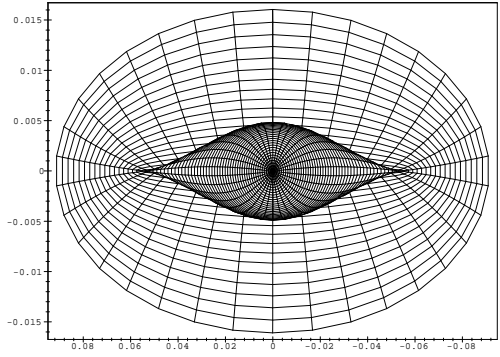}
        }
\caption{\label{fig:singlelip}The single lips regime ($T=T_l+9800$).
  The bottom panel shows a front view of the lips singularity.}
\end{figure}

At a later time, the first singular regime develops, illustrated
in Figure~\ref{fig:singlelip}. The tip of the spike, which earlier
was convex toward the observer, develops a self-intersection and a
sharp cuspidal ridge in the form of horizontal ``lips.''   Our use
of the term ``lips'' in this context is entirely for descriptive
purposes and should not be understood in the technical sense used
in Arnold's theory, in which the term ``lips'' refers to the $A_3
(+)$ planar caustic metamorphosis (e.g., pp. 375-376, 381 of \cite{PLW}).
The surface of the wavefront that faces the
observer is now concave, and limited by the two cuspidal ridges
outlining the ``lips.'' The two cuspidal ridges pinch off the back
convex sheet of the wavefront at two symmetrical swallowtail
points. The swallowtails are best appreciated in the top panel,
where a side view of the wavefront is shown. The bottom panel
shows a front view, where the lip character of the cuspidal ridge
can be appreciated. The wavefronts in this regime all have this
horizontal lip singularity, but the ``lips'' grow in time
continuously after starting out from a single point at the center.
The picture shows the wavefront at $T=T_l+9800$, which is still
quite early in this regime. One notable fact about this wavefront
is that, locally, it is indistinguishable from the early singular
wavefronts in the implosion of a triaxial ellipsoid
\cite{EFN01}, in which case, the cuspidal
lip-ridge lies ``in the outside.'' For the gravitational lensing
case that we are illustrating, in a global sense, the cuspidal
lip-ridge lies ``inside.''

\begin{figure}
\vbox{
\includegraphics[width=3in,angle=-90]{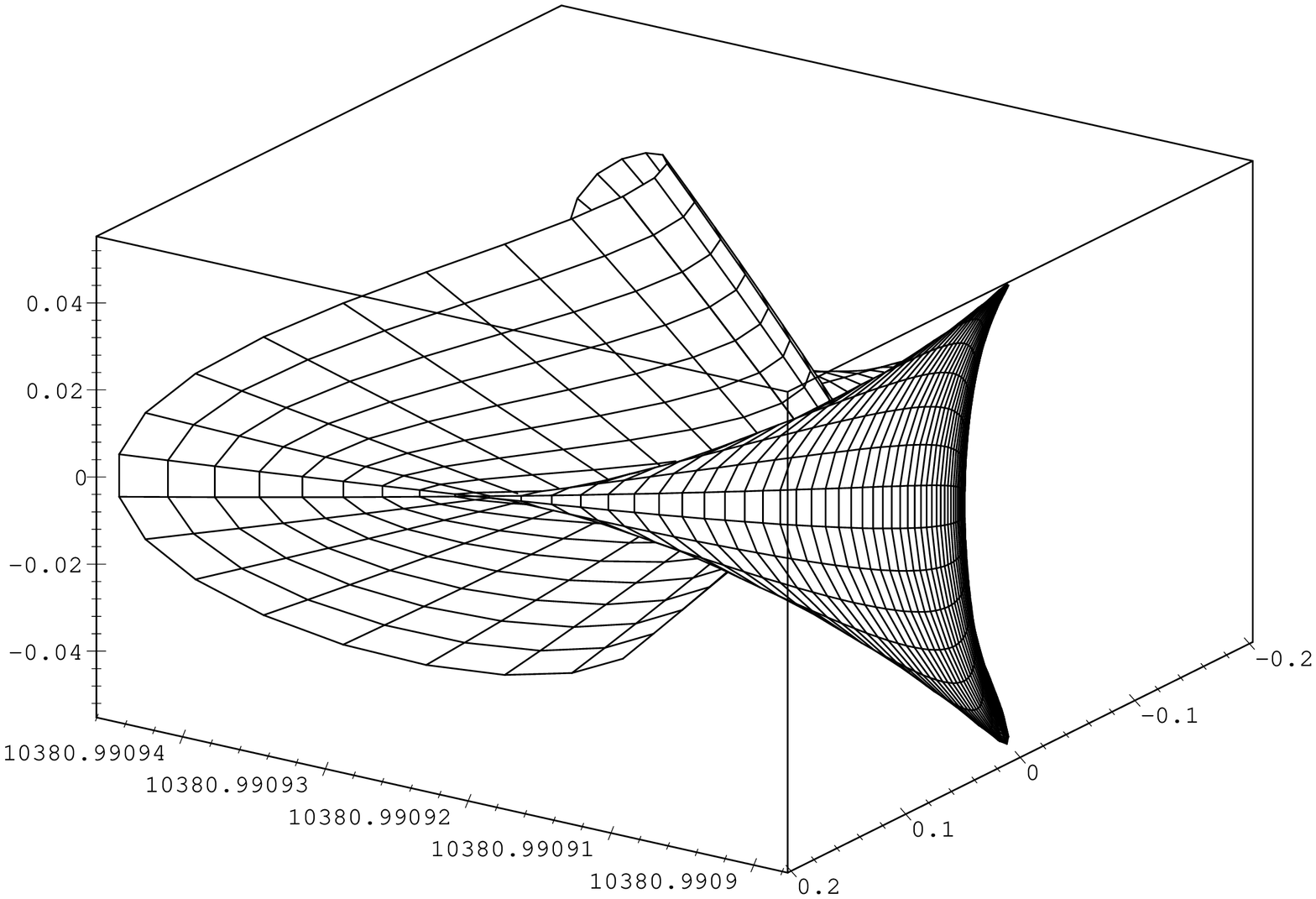}\\
\includegraphics[width=2.5in,angle=-90]{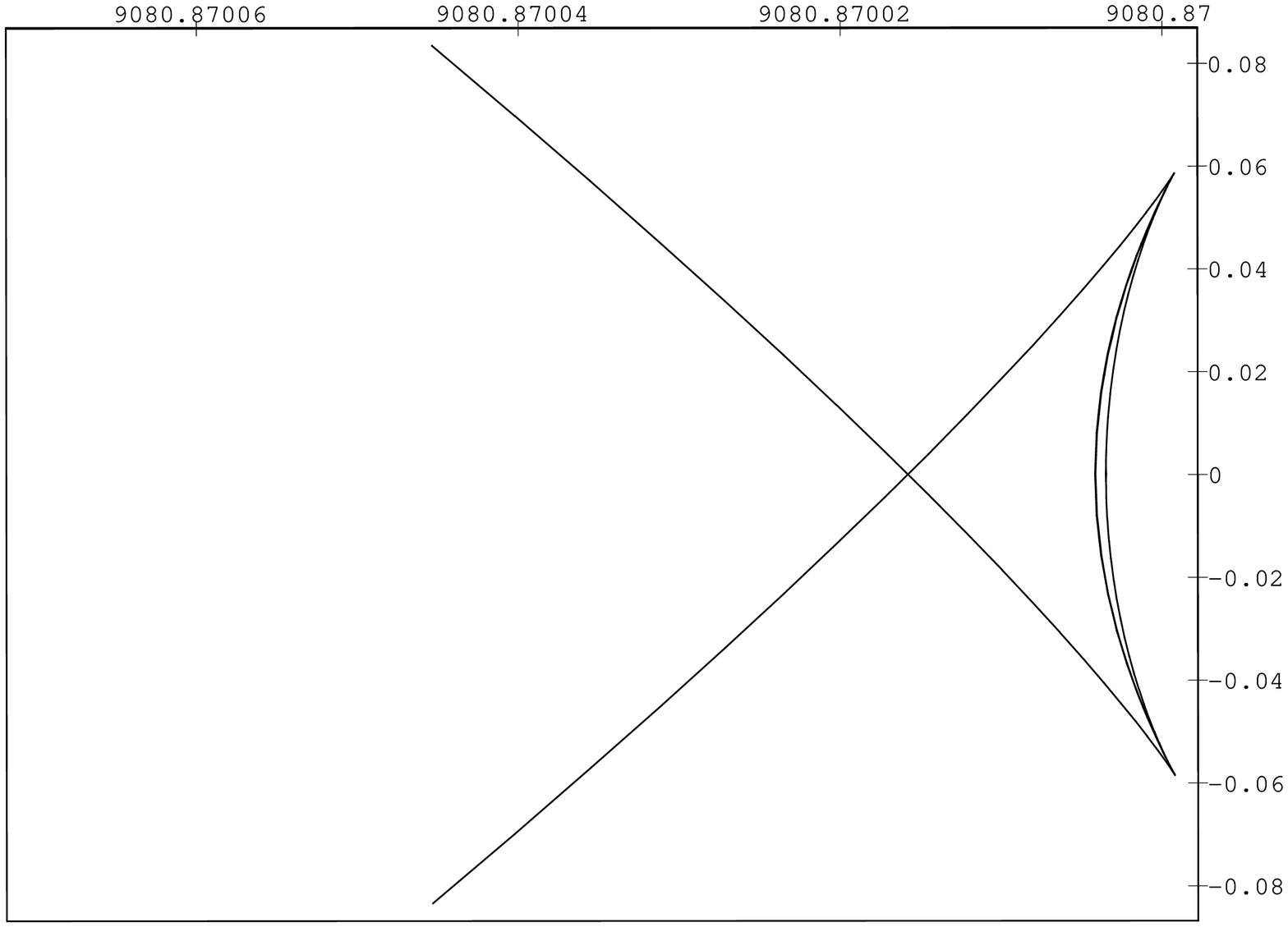}\\
\includegraphics[width=2.5in,angle=-90]{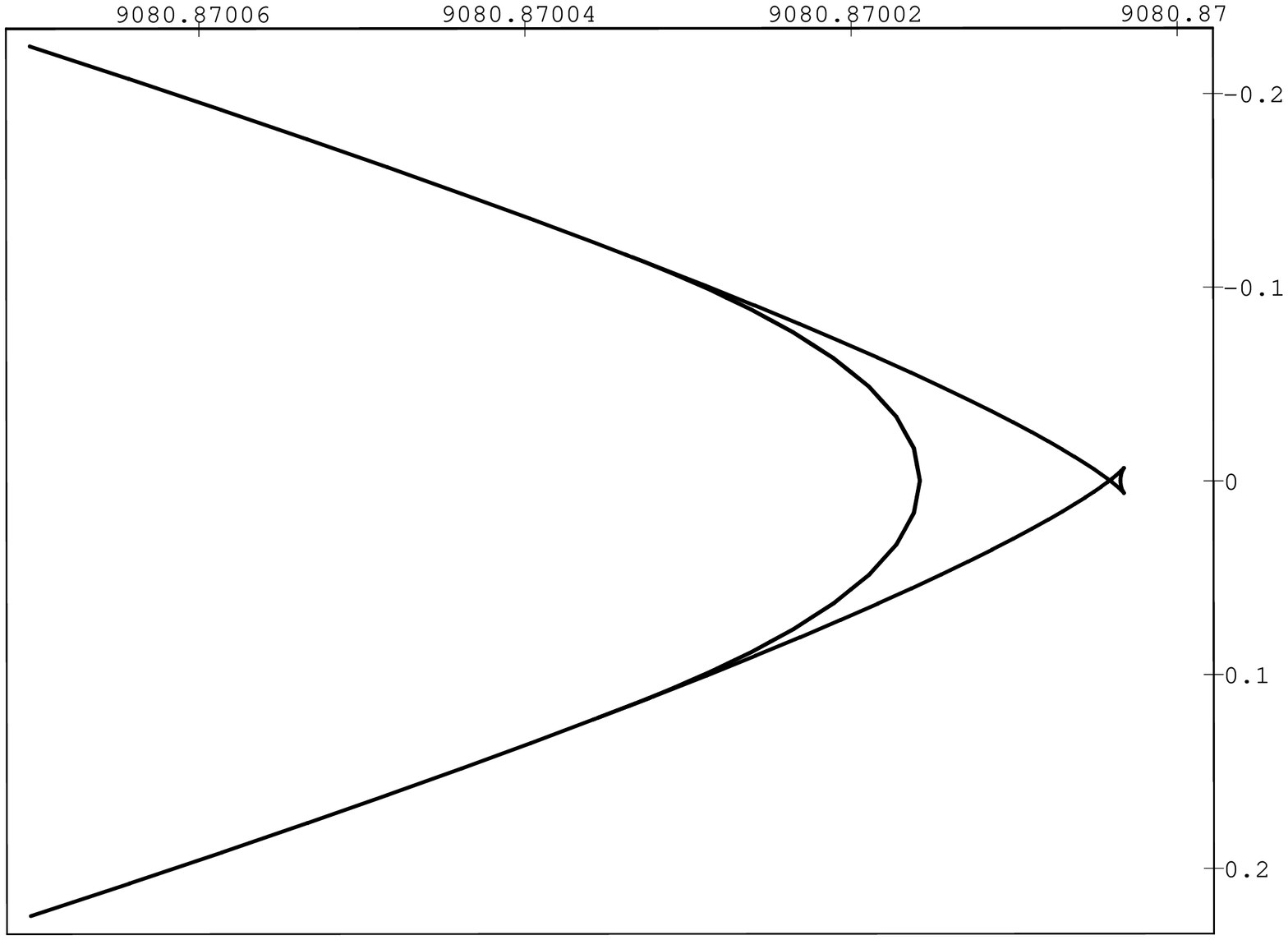}
        }
\caption{\label{fig:doublelip} Typical wavefront in the criss-cross
lip regime ($T=T_l+10200$), where two nested perpendicular cuspidal
ridges are present. The middle panel is a vertical slice of the
top panel, while the bottom panel is a horizontal slice.
Both slices are through the optical axis.}
\end{figure}

The next singular regime is illustrated in
Figure~\ref{fig:doublelip}. The point at the center of the concave
sheet of the wavefront facing the observer turns momentarily
singular at approximately $T=T_l+10148.84$, and grows another lips
singularity, this time vertically oriented. These vertical
``lips'' ride on top of the now mixed convex-concave section of
the wavefront and present a concave surface to the observer, as
can be appreciated from the two slices in the middle and bottom
panels. The middle panel shows a vertical slice of the wavefront
through the optical axis. It may be difficult to appreciate that
the foremost sheet pinches off the underlying originally existing
sheet well before touching the originally existing cuspidal ridges
at the top and bottom.  The bottom panel shows an optical-axis
horizontal slice, where the cuspidal character of the newly
created forefront sheet is evident.  The two ends of the vertical
``lips'' are two symmetrical swallowtails.  All the wavefronts in
this regime have two perpendicular lips, yielding a total of four
swallowtails.  The perpendicular lips grow continuously in time,
but the inner lip starts at the center and grows until it touches
the outer lips. This picture shows the wavefront at $T=10200$,
which is quite late within this regime since the inner lip is
about to merge with the outer one.

\begin{figure}
\vbox{
\includegraphics[width=3in,angle=-90]{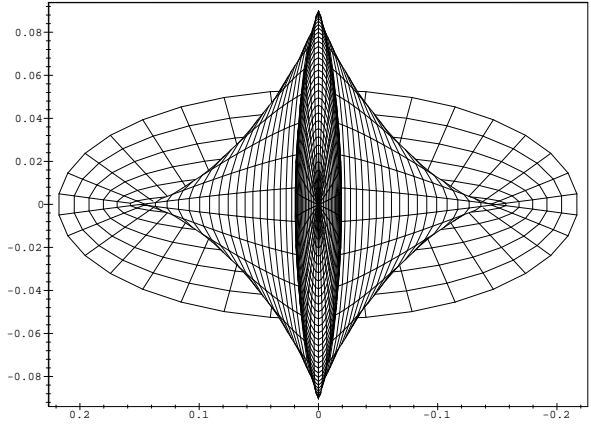}
        }
\caption{\label{fig:football} The critical wavefront ($T=T_l+10420$).}
\end{figure}

The next regime consists entirely of a single wavefront at the
critical time when the inner lip singularity merges with the outer
one. This wavefront is represented in Figure~\ref{fig:football}.
The singular ridge on the wavefront has the general outline of a
football in a vertical position, inscribed in an outer astroid or
diamond. The ``football'' outline is the distorted last expression
of the inner lip-ridge, whereas the astroid is the distorted last
expression of the outer (original) horizontal lips. At this time
the two cuspidal ridges merge right before they undergo a
permanent change. The wavefront at this critical time has two
swallowtails symmetrically located along the horizontal direction,
and two singular points symmetrically located along the vertical
direction where the two lips merge. In the context of wavefronts,
these points have no greater significance than the merger of a
cusp ridge and a swallowtail surface. However, their greater
significance is that they are indicators of the presence of
hyperbolic umbilic points in the caustic sheet, which we describe
in the next subsubsection. The figure represents the time
$T=10420$, which is close enough to the critical time at the
resolution that we are using.

\begin{figure}
\vbox{
\includegraphics[width=3in,angle=-90]{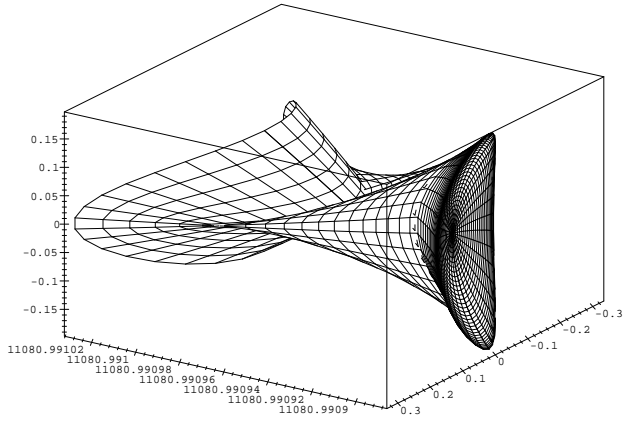}\\
\includegraphics[width=3in,angle=-90]{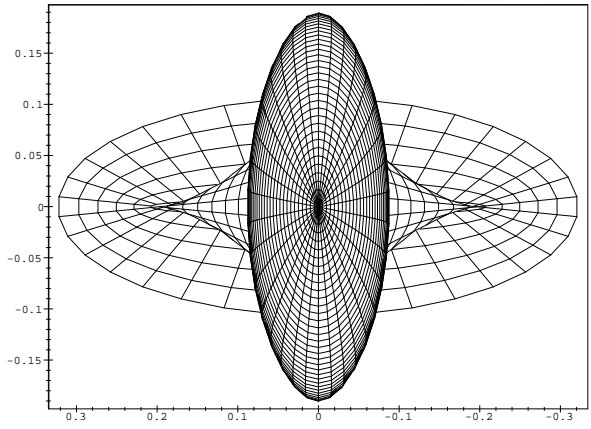}
                }
\caption{\label{fig:astroid} Typical wavefront early in the
late regime ($T=T_l+10900$). The bottom panel shows a front
view.}
\end{figure}

\begin{figure}
\vbox{
\includegraphics[width=1.79in,angle=-90]{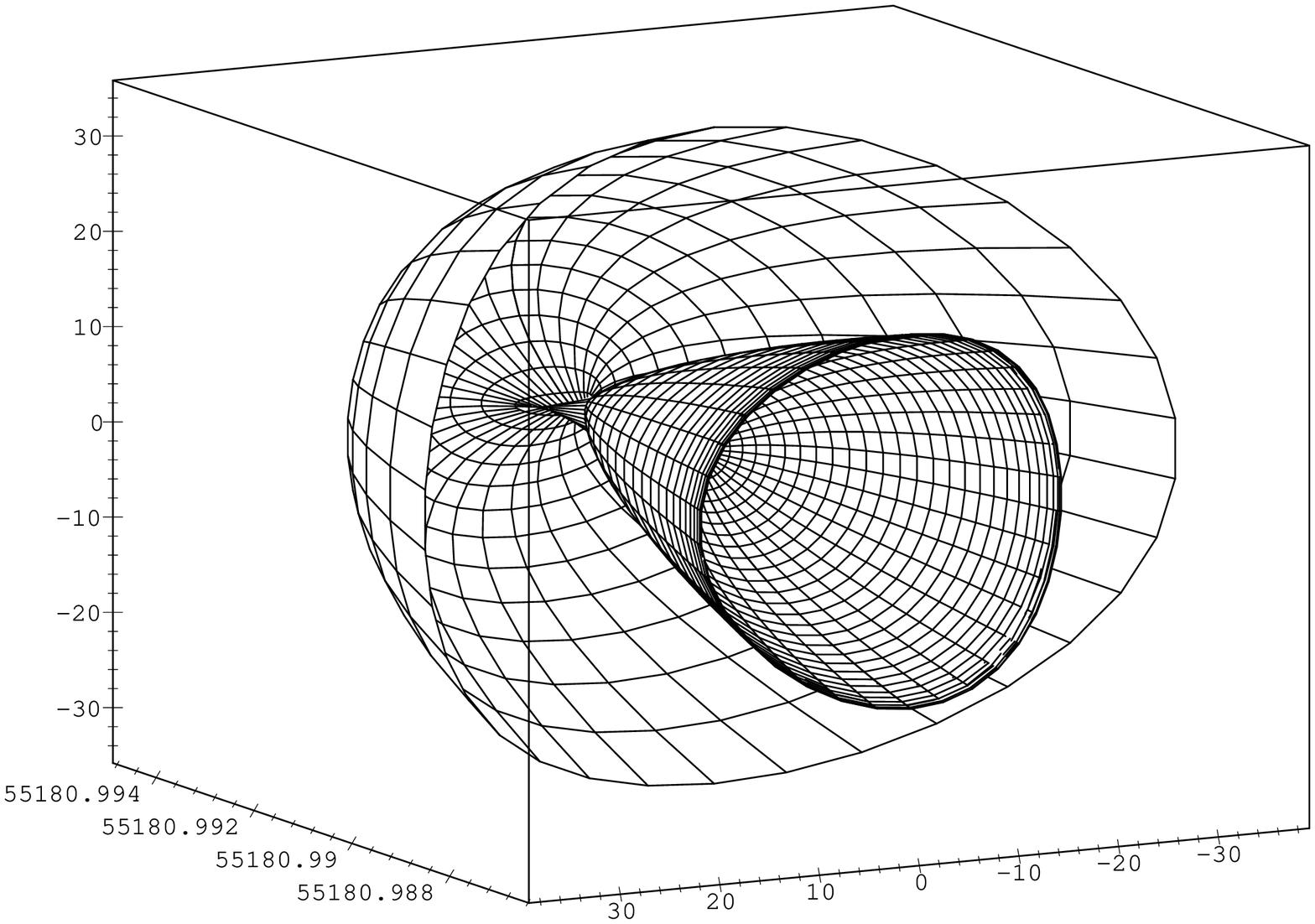}\\
\includegraphics[width=1.79in,angle=-90]{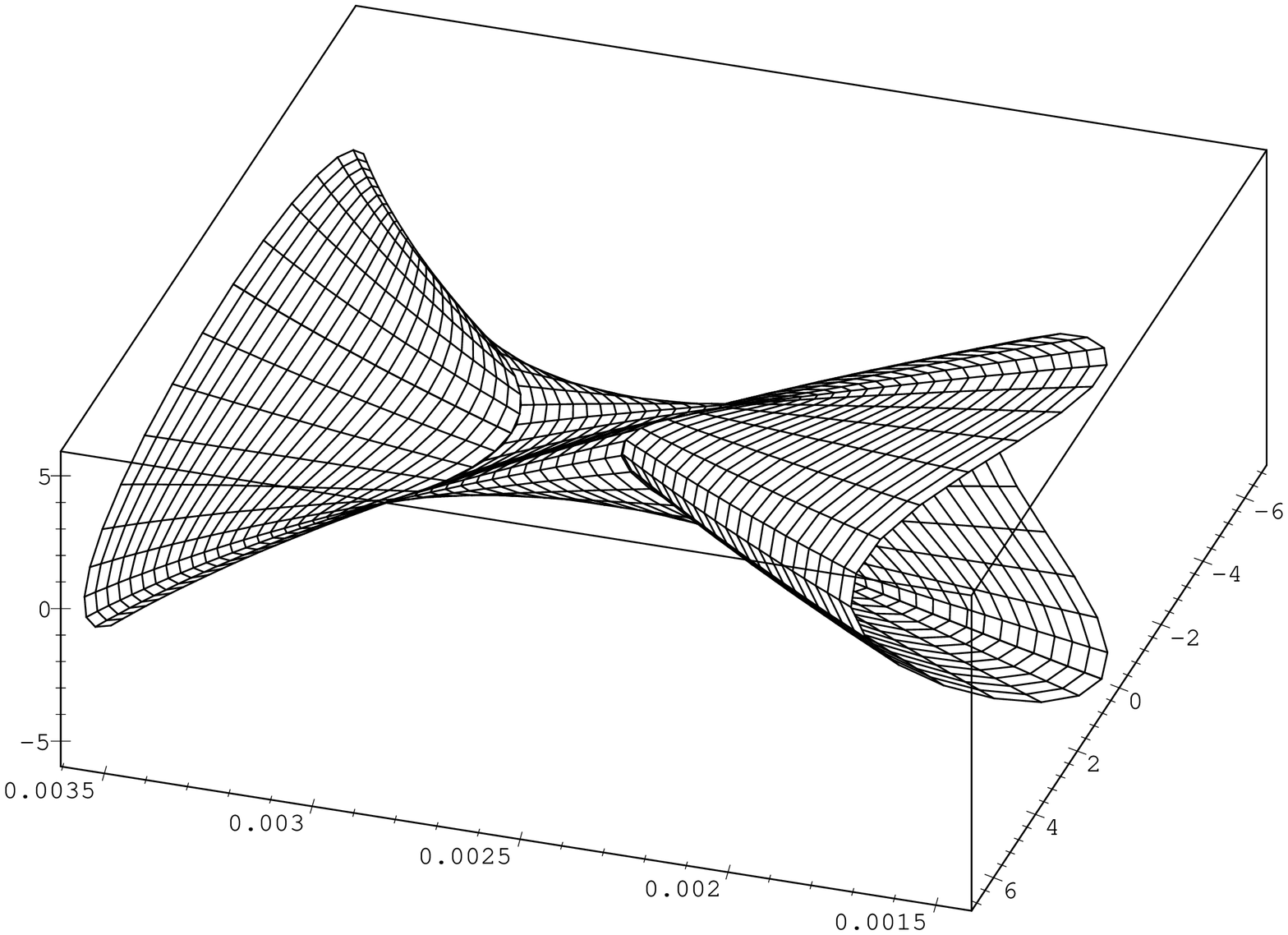}\\
\includegraphics[width=1.79in,angle=-90]{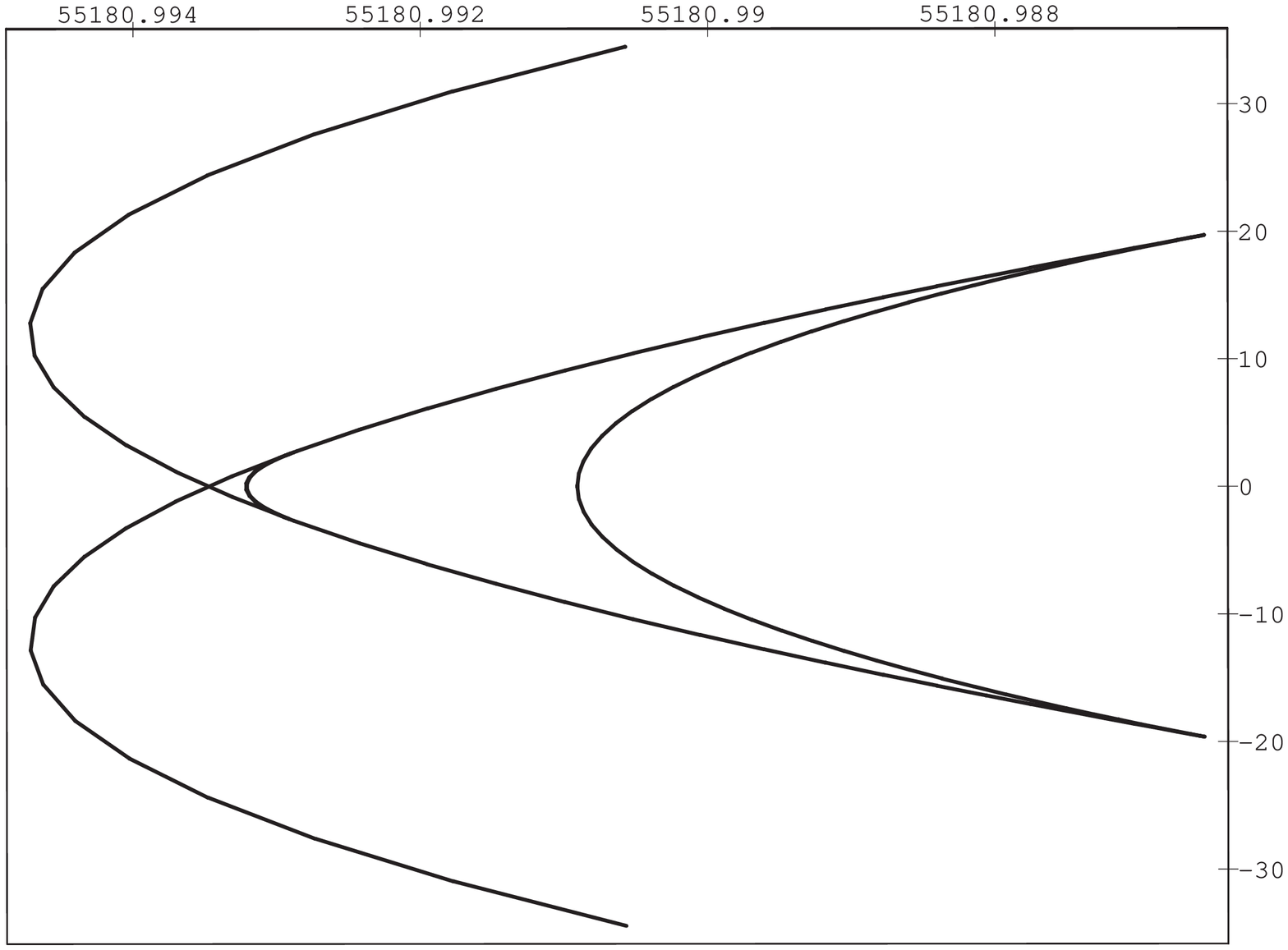}\\
\includegraphics[width=1.79in,angle=-90]{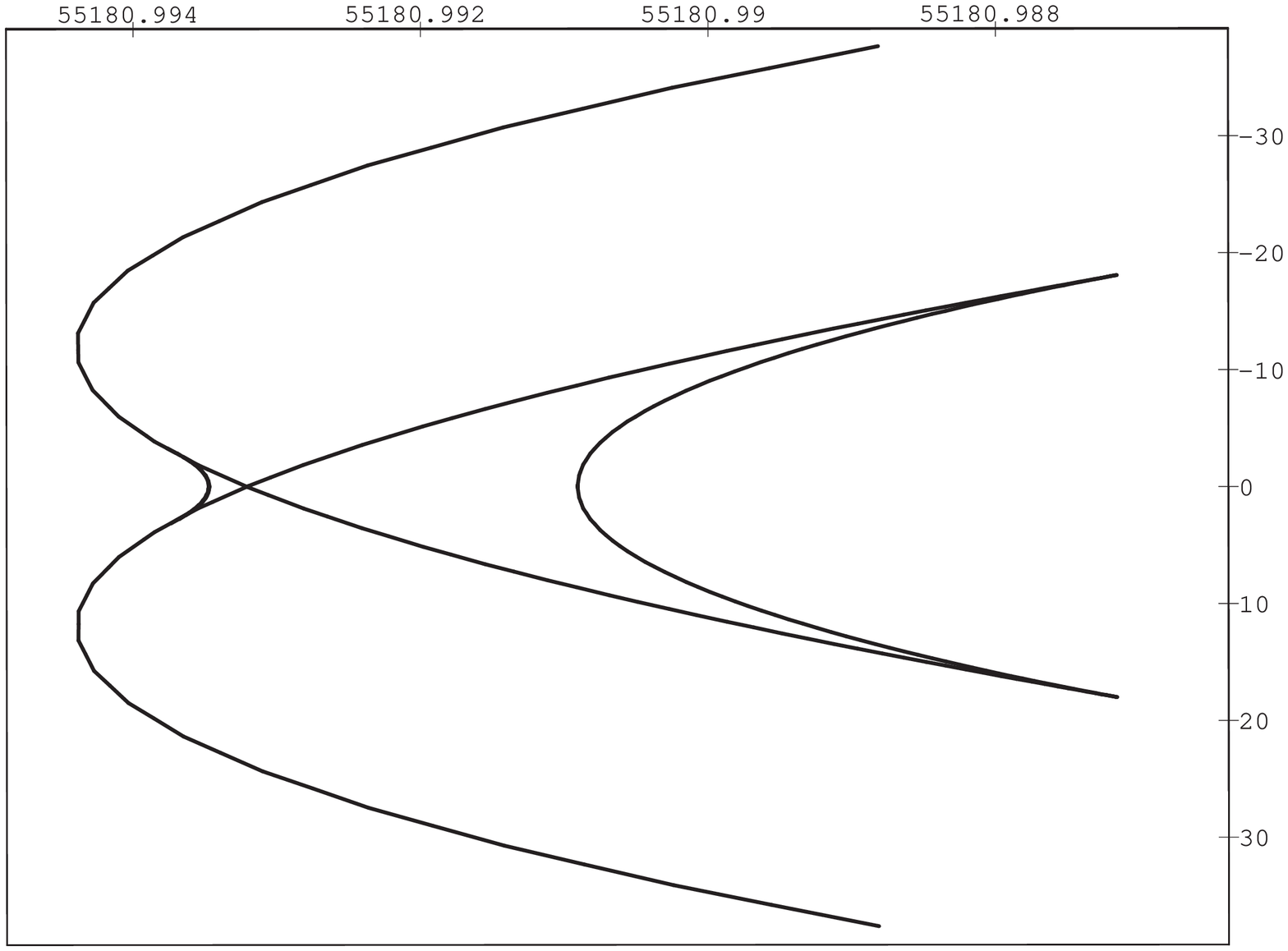}
                }
\caption{\label{fig:goblet} Typical ``goblet'' wavefront in the very
late regime (in this case, $T=T_l+55000$). The middle-top panels shows a
magnification of the throat. The middle-bottom and bottom panels show vertical
and horizontal slices, respectively.}
\end{figure}

The last regime comprises all the wavefronts later than the
critical time at which the inner ``lips'' merge with the
outer    ``lips.''   All such late wavefronts have a ``regular''
cuspidal ridge delimiting a concave surface that faces the
observer, namely, a cuspidal ridge which is smooth as a curve
in three space. Behind the foremost concave cap, the wavefront
self-intersects and has another cuspidal ridge with four
singular points in the general shape of an astroid or diamond.
One such wavefront at time $T=T_l+10900$ is shown in
Figure~\ref{fig:astroid}. In this front view, the foremost oval
cuspidal ridge is clearly distinguishable, as are the two
swallowtails in the back.  However, there are two other
swallowtails completing the diamond, which lie behind the
foremost oval cap and are blocked from view in this picture.
The qualitative structure of the wavefront in the diamond-ridge
neighborhood is shown in the bottom panel of
Figure~\ref{fig:goblet}.

A very late wavefront at time $T=T_l+ 55000$ is shown in
Figure~\ref{fig:goblet}.  It can be seen that as time moves on,
the oval and diamond cuspidal ridges remain essentially
unchanged except that the oval ridge in the foreground grows at
a much faster rate than the diamond-shaped ridge in the back,
and the wavefront eventually acquires a shape resembling that
of a goblet. The figure shows both a field view of the
``goblet'' and a magnification of the goblet's throat to show
the diamond ridge. The diamond ridge structure is almost
ubiquitous in wavefront evolutions that are not axi-symmetric.
For another view of one of these local diamond ridges in
wavefronts, we refer the reader to the bottom panel of
Figure~\ref{fig:goblet0}.  The global ``goblet'' wavefront is
typical of spherically symmetric regular potentials, with the
major difference that the throat of the goblets in the
spherically symmetric case degenerates down to a single point,
lacking the complicated diamond structure for the case
of elliptical symmetry.    The diamond ridge
represents the nondegenerate version of the throat of the
goblet. The goblet's throat has three wavefront
sheets, each generically giving rise to an image of the
light source.

What we have described in this section are precisely the
singularities and metamorphosis of wavefronts in the observer's
lightcone in the case of an elliptical potential. In this case, we
found that the wavefronts have three types of singularities:
cuspidal ridges, swallowtail points, and points of transversal
self-intersections.  These three singularity types are shared by
generic wavefronts in space --- see p. 55 of \cite{Arn91}.  We
also showed  that two kinds of metamorphosis occur for the
wavefronts due to an elliptical potential. In the first place, we
have found two occurrences of the birth of ``lips'' on the
wavefront. Secondly, we have found two hyperbolic umbilic
metamorphoses, i.e, the two symmetrical occurrences on the
wavefront of the exchange of a swallowtail point from one cuspidal
ridge to another. Both metamorphosis are part of Arnold's list of
metamorphoses of fronts in space --- see, for instance, the first
and fourth perestroikas in Figure~258  on page 489 of \cite{Arn89}.

\subsubsection{Caustic Sheet \label{sec:regularcausticsheet}}

If we think of the family of wavefronts for all times $T$ as
spatial slices of the observer's lightcone, then the collection of
all the caustic singularities (i.e., cuspidal ridges, swallowtail
points) on the instantaneous wavefronts forms a two-surface in
spacetime.  The projection of this two-surface into the comoving
space $E$ is referred to as a comoving caustic surface.

As a surface in the three-space $E$, the caustic surface is traced
in time by the caustics of the traveling wavefront as the front
moves away from the observer. The metamorphoses of the wavefront's
caustics thus build up a picture of the caustic surface's
singularities. The generic singularities of a caustic surface in
three-space are of five different types: folds, cuspidal ridges,
swallowtail points, elliptic umbilic points and hyperbolic umbilic
points \cite{Arn86}.

\begin{figure}
\vbox{
\includegraphics[width=3in,angle=-90]{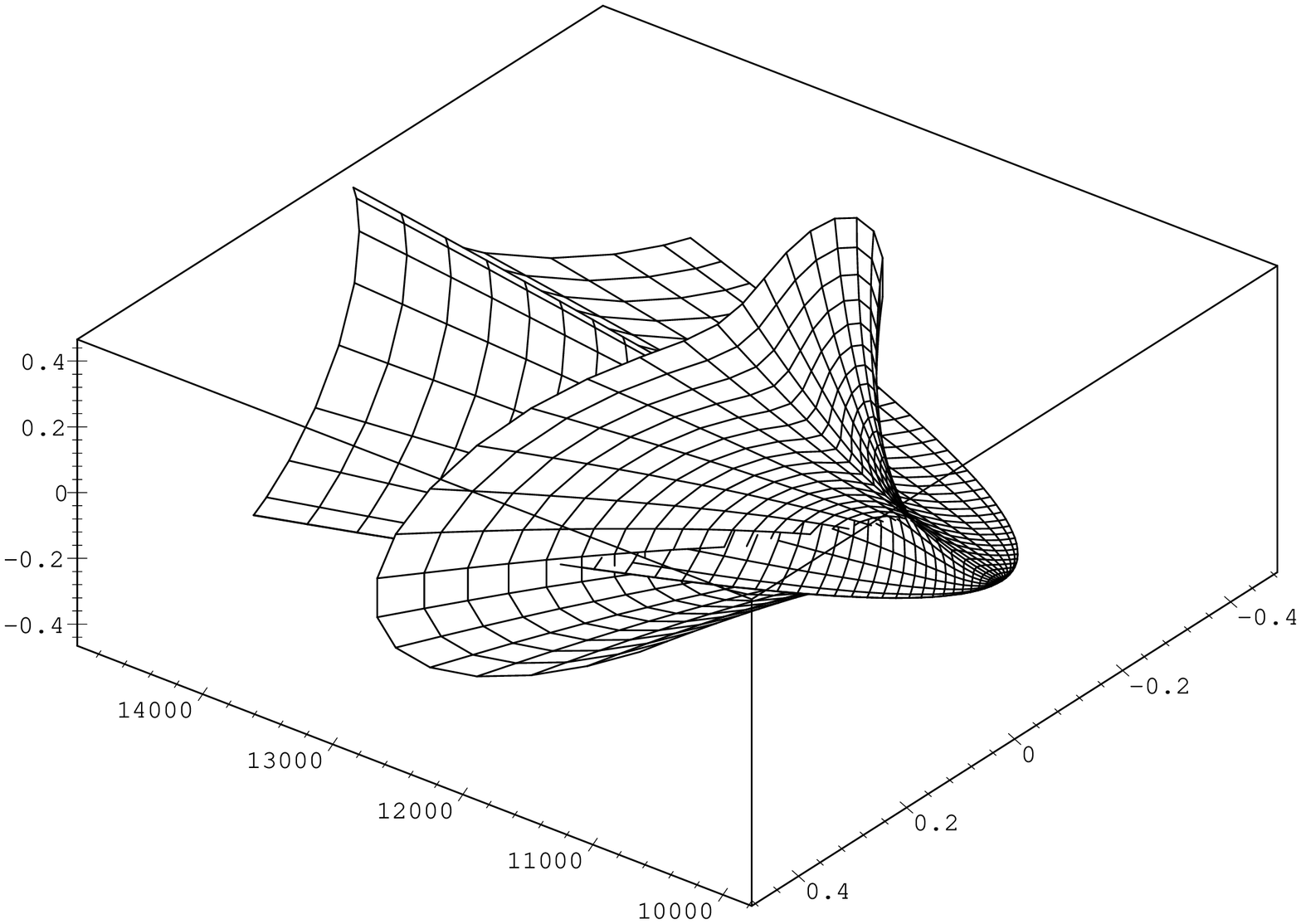}\\
\includegraphics[width=3in,angle=-90]{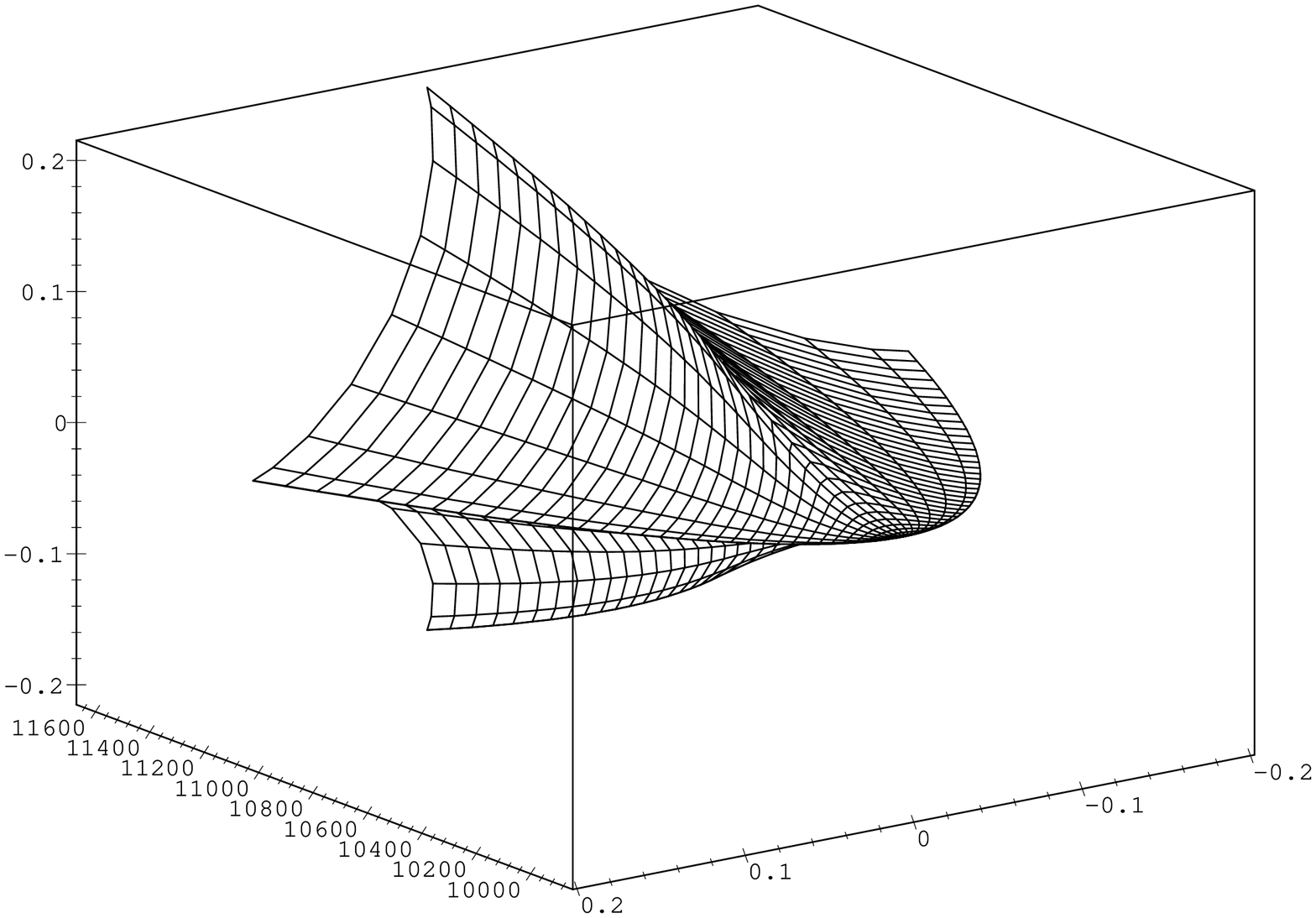}\\
\includegraphics[width=2.5in,angle=-90]{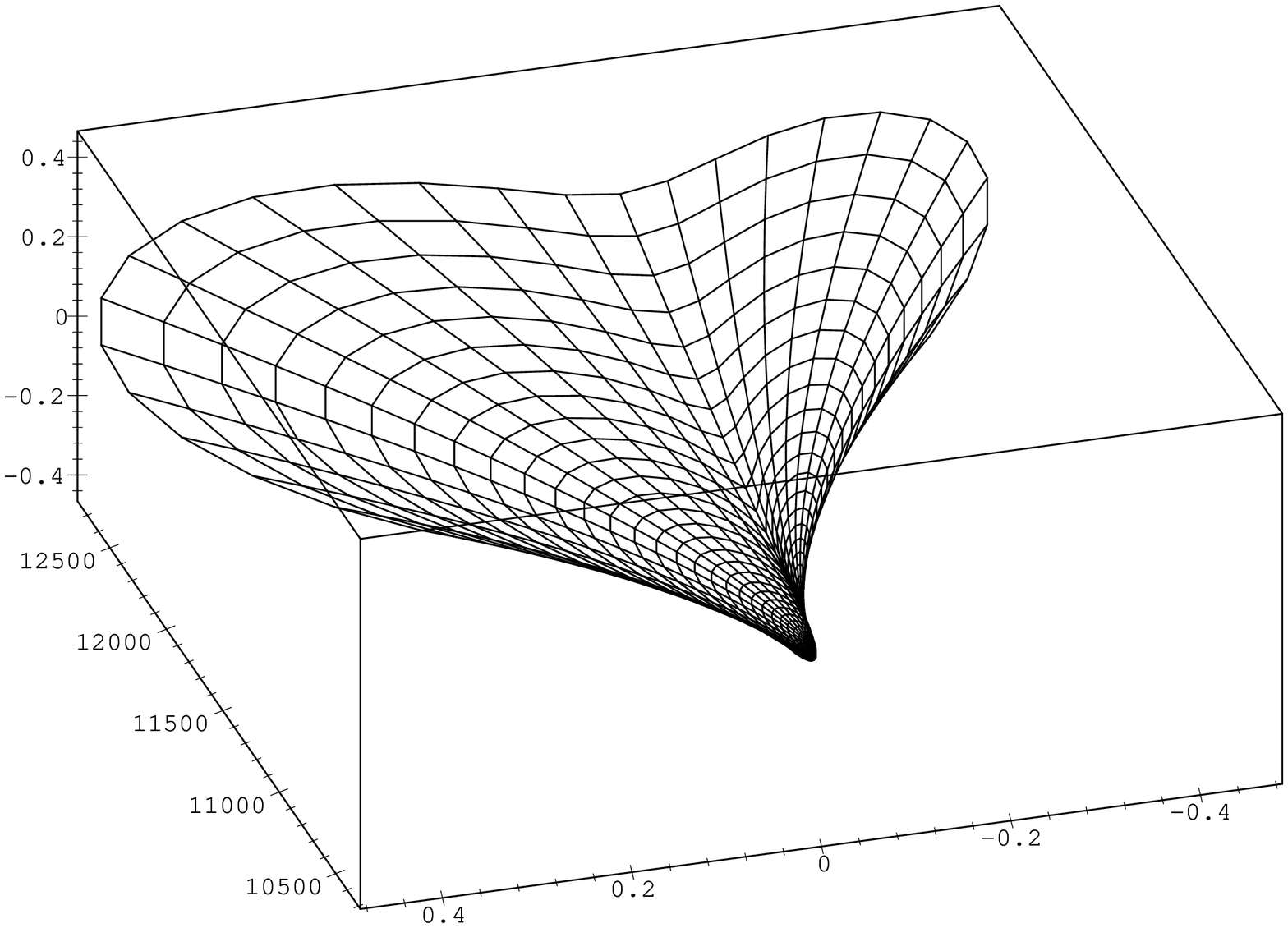}
        }
\caption{\label{fig:caustic}
Caustic sheet of the elliptical
potential. The two component sheets are represented separately in
the middle and bottom panels. }
\end{figure}

We showed that the caustic surface on an observer's lightcone is given by
equation (\ref{eq-comov-caustic-surface-z}). In the particular case of the
elliptical potential, the resulting caustic surface is shown in
Figure~\ref{fig:caustic}. A significant feature of our caustic surface in
three-space is that it consists of two separate intersecting sheets, one for
each non-trivial root (\ref{eq-zpm}) of the Jacobian determinant in
(\ref{eq-det-dszdr}). The two bottom panels in the figure show the two
component sheets. One sheet starts closer to the observer as a horizontal
``beak'' and then develops a diamond cross section.  The second sheet starts
inside the first one as a vertical ``beak'' and eventually opens up into a
smooth surface with an oval cross section.  The transitions of both sheets take
place at the points where they intersect. The points of intersection where the
vertical beak disappears from the second component sheet and the vertical cusp
ridge appears on the first component sheet are hyperbolic umbilic points  ---
there are two such points.

We can also at this point give the exact value of the time of
the first singular wavefront. This is the smaller of
$T(\bs_+(\bmo),z_+ (\bmo))$
and
$T(\bs_- (\bmo),z_- (\bmo))$ since
rays passing through the origin have the
longest delay (as is evidenced from the spike).
For the previous subsubsection, we have
$T(\bs_+ (\bmo),z_+ (\bmo))= T_l+9699.27$
and
$T(\bs_- (\bmo),z_-  (\bmo))=
T_l+10148.84$, i.e.,  the first singular wavefront is the
surface at $T=T_l+9699.27$, as anticipated.

\begin{figure}
\vbox{
\includegraphics[width=1.79in,angle=-90]{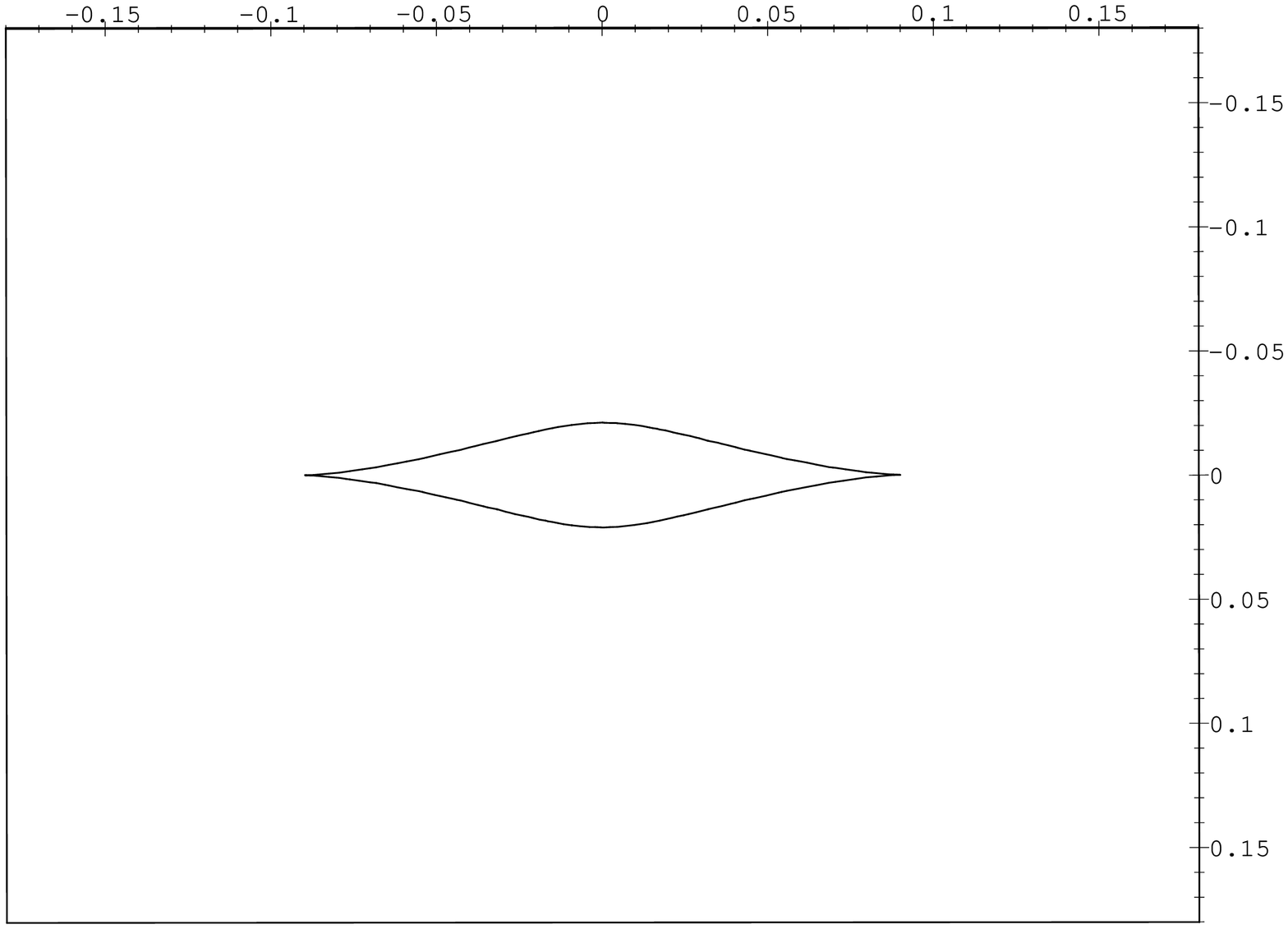}\\
\includegraphics[width=1.79in,angle=-90]{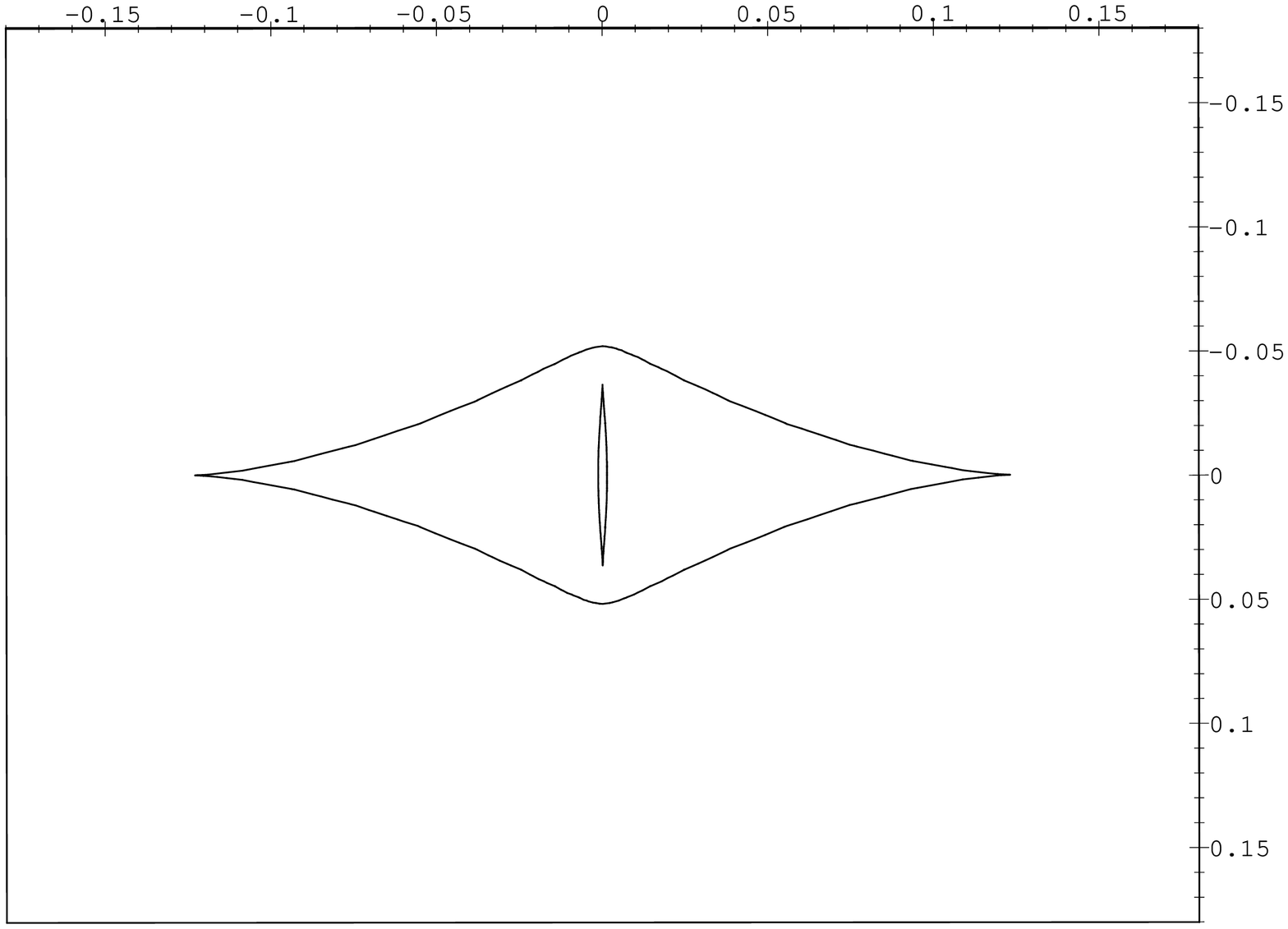}\\
\includegraphics[width=1.79in,angle=-90]{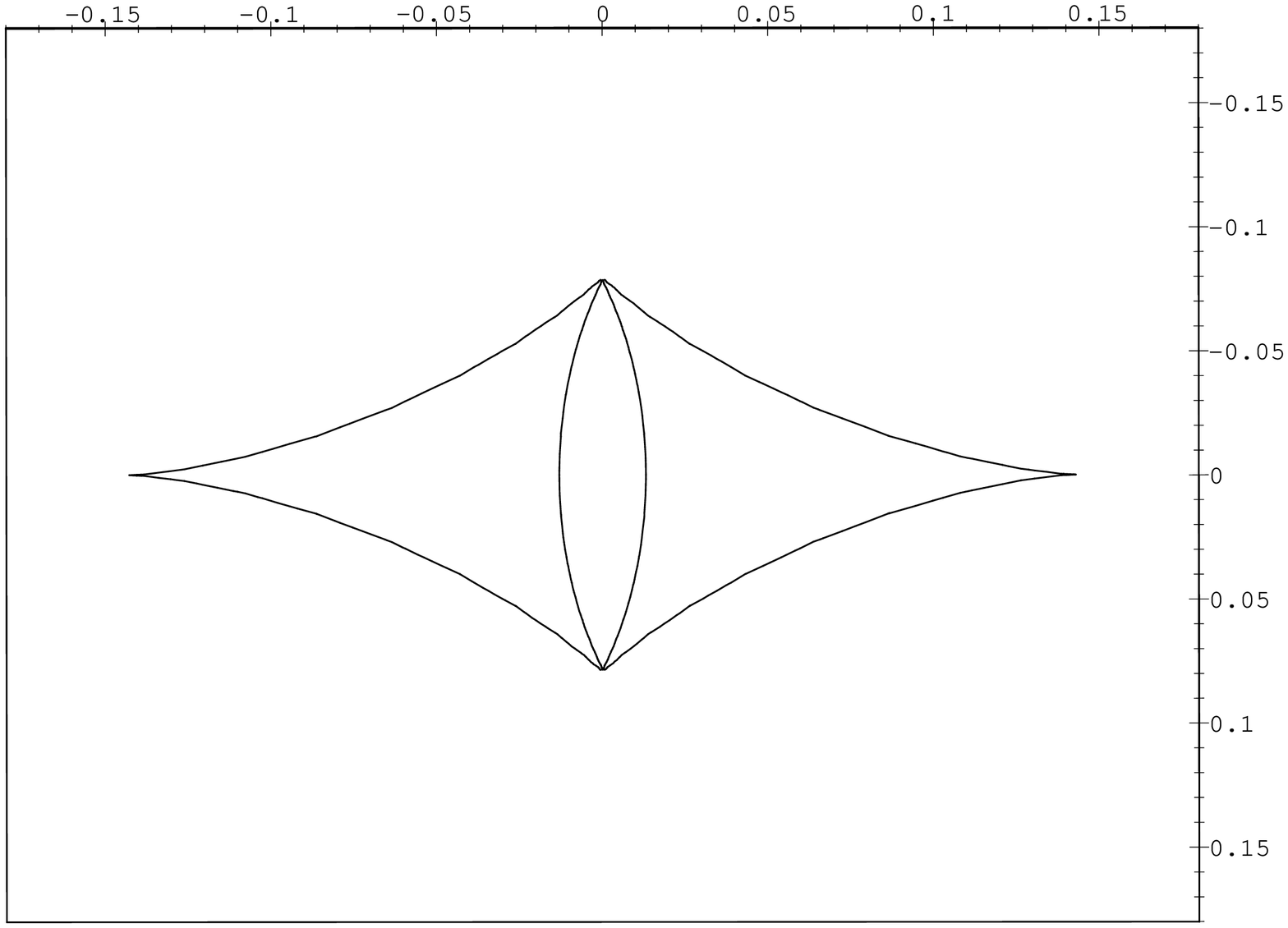}\\
\includegraphics[width=1.79in,angle=-90]{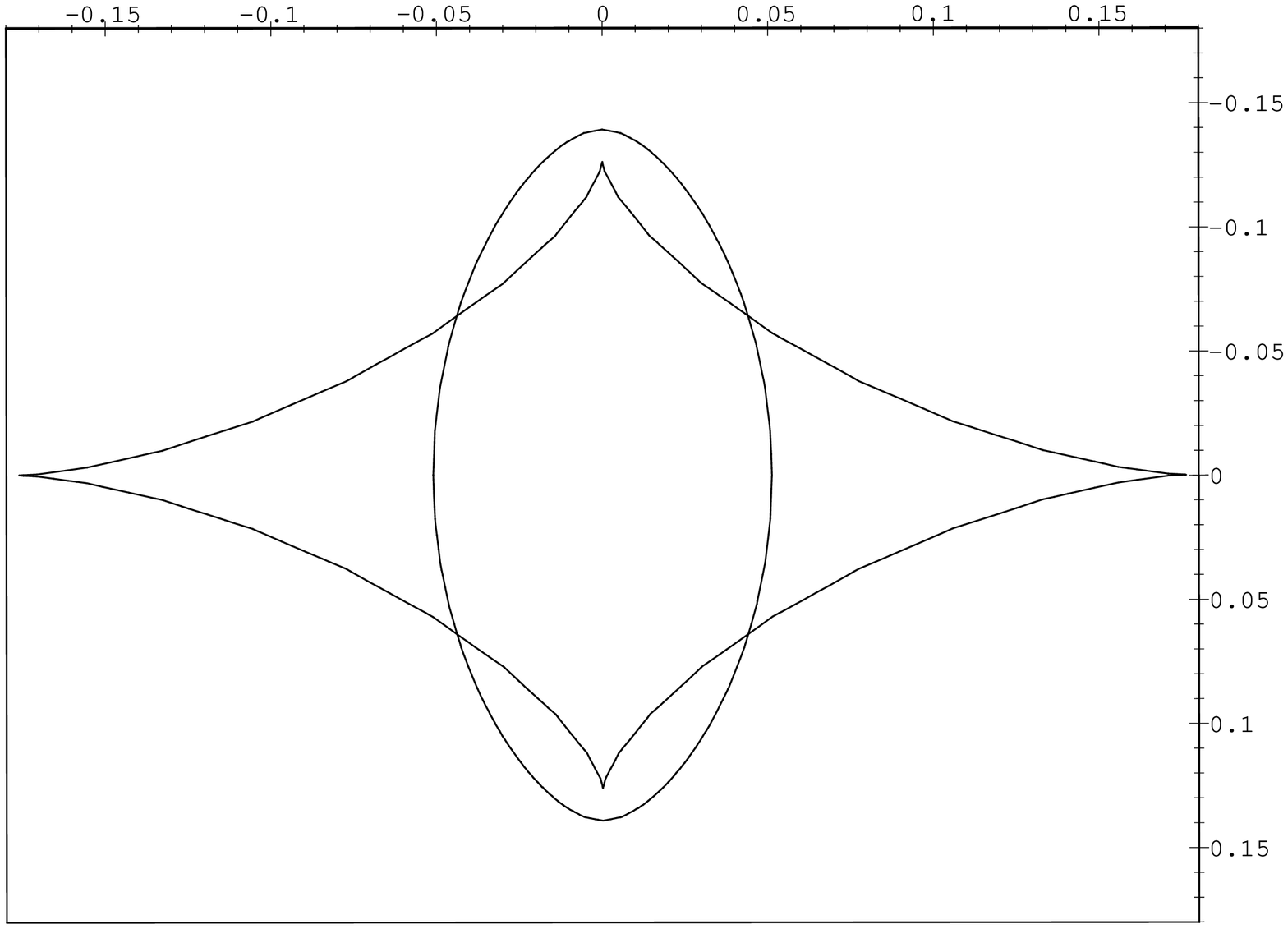}\\
\includegraphics[width=1.79in,angle=-90]{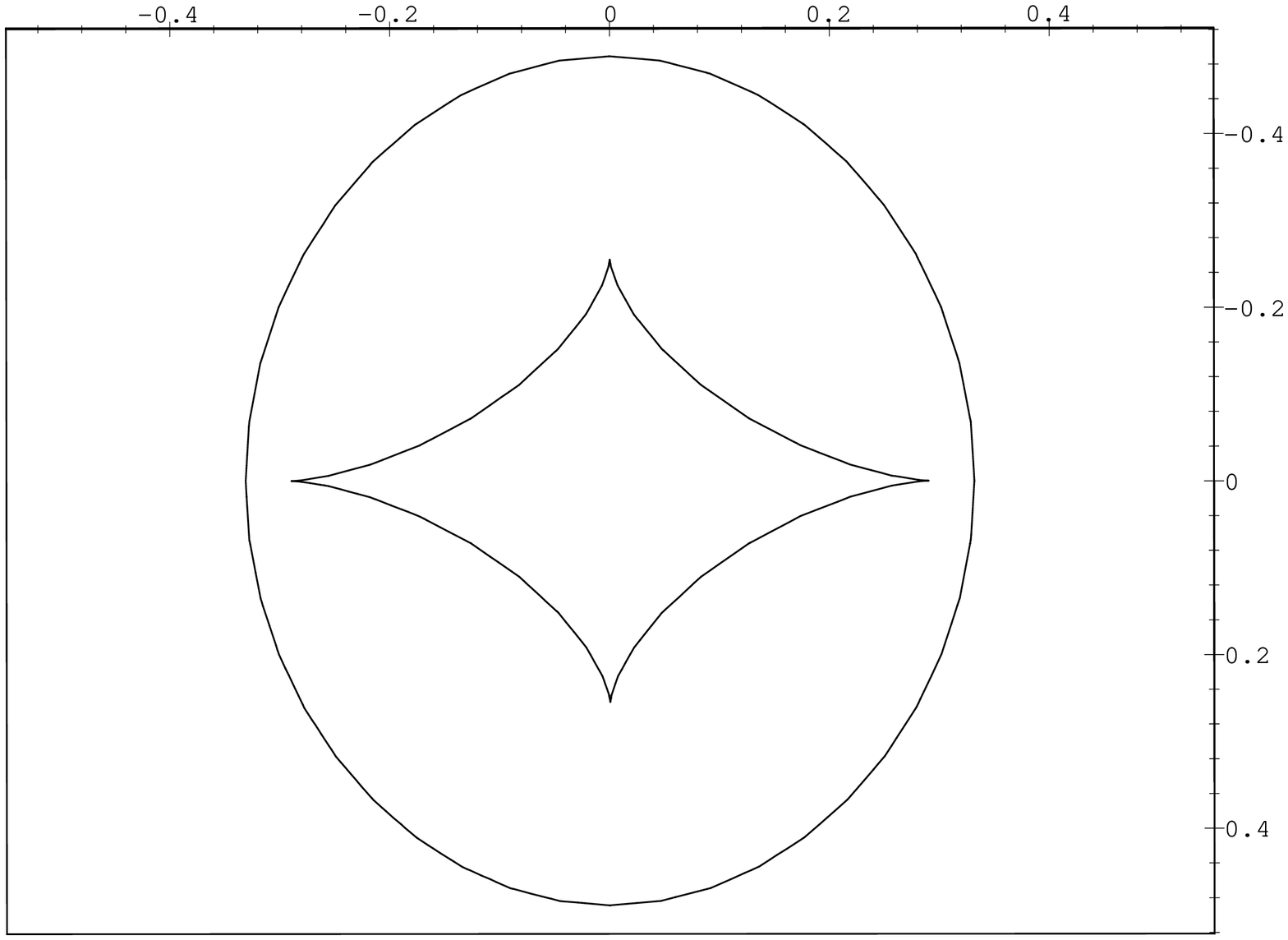}
        }
\caption{\label{fig:planarcaustic} The metamorphosis of the planar
caustics arising from $z$-plane slices of the caustic surface.}
\end{figure}

By slicing the caustic sheet with constant-$z$ surfaces we obtain
a series of planar caustics representing the metamorphosis of the
hyperbolic umbilic.  These are shown in
Figure~\ref{fig:planarcaustic} and coincide with the well-known
caustic curves in the gravitational lensing literature for an
elliptical potential
--- e.g., p. 386 of \cite{PLW}.
Beautiful photographs of caustic metamorphoses are
shown in \cite{Ny99}  --- see p. 68 for the hyperbolic
umbilic metamorphosis.

\subsection{Singular Elliptical Potential Lens \label{subsec:sing-ep}}

We shall illustrate the wavefront singularities and
caustic surface due to a singular elliptical potential,
namely, the case of vanishing core radius ($r_c =0$).
In this case, it is
simpler to use polar coordinates  $(r, \vartheta)$
on the lens plane. We have for
the potential,
\beqan
    \hp_{ep}^{sing} (r,\vartheta) 
&\equiv&
           A_0
        \sqrt{(1-\epsilon)r_1^2+(1+\epsilon)r_2^2}
\\
&=&  A_0 \  r
     \sqrt{1+\epsilon\cos(2\vartheta)},
\eeqan
and for the associated bending angle,
$$
\hat{\bma}_{ep}^{sing} (r, \vartheta)
   =    \frac{A_0}
         {\sqrt{1+\epsilon\cos(2\vartheta)}}
    \Big( (1\!+\!\epsilon) \cos\vartheta, (1\!-\epsilon)\sin\vartheta\Big),
$$
which has no dependence on $r$. The potential vanishes at the
origin $r=0$. This case is singular because the surface mass
density associated with the potential blows up at the origin of
the lens plane (i.e.,  $\nabla^2_\br \hp_{ep}^{sing} (\br)
\rightarrow \infty$  as $\br \rightarrow \bmo$).    The center of
the lens (i.e., the origin)  thus acts as an obstruction to the
light rays --- see \cite{Ptt95}) and p. 544 of \cite{PLW}
for a detailed treatment of obstruction points. Soon after
going through the lens plane, the wavefronts evolve markedly
differently from the nonsingular case. For our illustrations, we
have used the following values of the lens parameters: $A_0 = 0.3,
\epsilon=0.02, z_l=4000$. As in the regular case, these values are
chosen for pedagogical reasons, although taking care to respect
the approximations of small angles.

\subsubsection{Wavefront Singularities}

\begin{figure}
\vbox{
\includegraphics[width=3in,angle=-90]{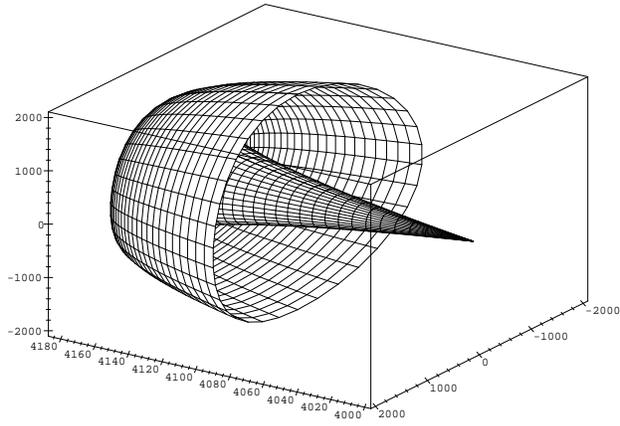}\\
\includegraphics[width=2.5in,angle=-90]{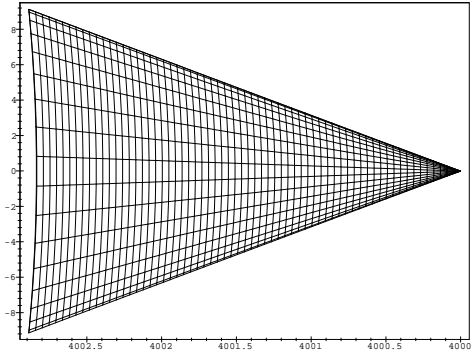}
        }
\caption{\label{fig:spike0} The unique wavefront in the early
regime ($T=T_l$) in the singular case.  The bottom panel shows a
side view of the tip of the spike.}
\end{figure}

\begin{figure}
\vbox{
\includegraphics[width=3in,angle=-90]{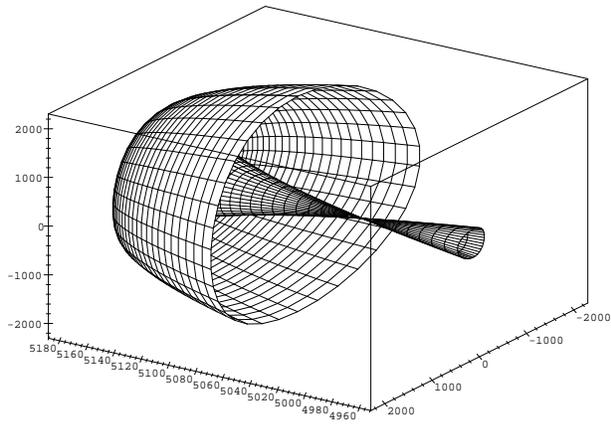}\\
\includegraphics[width=2.5in,angle=-90]{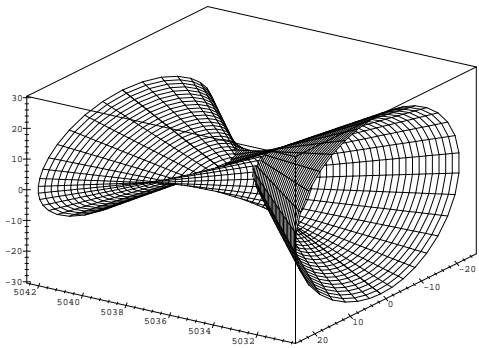}
        }
\caption{\label{fig:goblet0} Typical wavefront in the late regime
($T=T_l+1000$) in the singular case. The bottom panel shows a
magnified view of the ``throat'' of the goblet.}
\end{figure}

Due to the obstruction at the center of the lens (i.e., the origin),
all the wavefronts are singular and there are only two regimes.
The first regime consists entirely of a single wavefront
(the
earliest one to make it past the lens plane)  with
an interior spike pointing towards the observer,
as did the wavefronts of the nonsingular elliptical
potential. However, in this singular case  the spike is
conical, and touches the lens plane at the origin. The point at
the tip of the spike is removed. This wavefront, $T=T_l=4000$,
is shown in Figure~\ref{fig:spike0}. The figure shows both a
field view of the wavefront, and a side view of the tip of the
spike magnified to make the conical character apparent.

At all times afterwards, the wavefronts resemble a goblet --- the second regime.
However, these goblets are markedly different from those
that resulted at late times in the nonsingular elliptical potential
case. There is a
single diamond-shaped cuspidal ridge, which lies at the throat
of the goblet.   The rim of the goblet is removed and acts
as a boundary of the wavefront. These goblets lack the interior
concave surface facing the observer that characterizes the
goblets in the nonsingular case.
The throat of the goblet has three sheets, each typically giving
rise to an image of the source.  One such wavefront in this regime
($T=T_l+1000$) is illustrated in Figure~\ref{fig:goblet0},
where a field view evidences the resemblance to a goblet.  The
bottom panel in the figure shows a magnified view of the throat
of the ``goblet'' where the diamond structure is evident.

\begin{figure}
\includegraphics[width=3in,angle=-90]{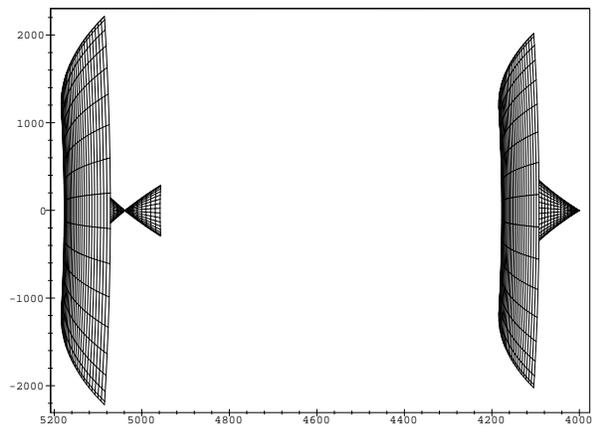}
\caption{\label{fig:history0} Progression of a portion of the
wavefront in the singular case. The observer is on the right. }
\end{figure}

Two questions relating to this imperfect ``goblet'' wavefront
arise. In the first place, do the wavefronts ever detach from
the lens plane, or, on the contrary, do they reach out to the
obstruction point?  As one can appreciate in
Figure~\ref{fig:history0}, where a given portion of the initial
wavefront at $T=T_l$ is followed up and plotted again at
$T=T_l+1000$,  {\it the wavefront does not remain attached to the
lens plane.}

\begin{figure}
\includegraphics[width=2.5in,angle=-90]{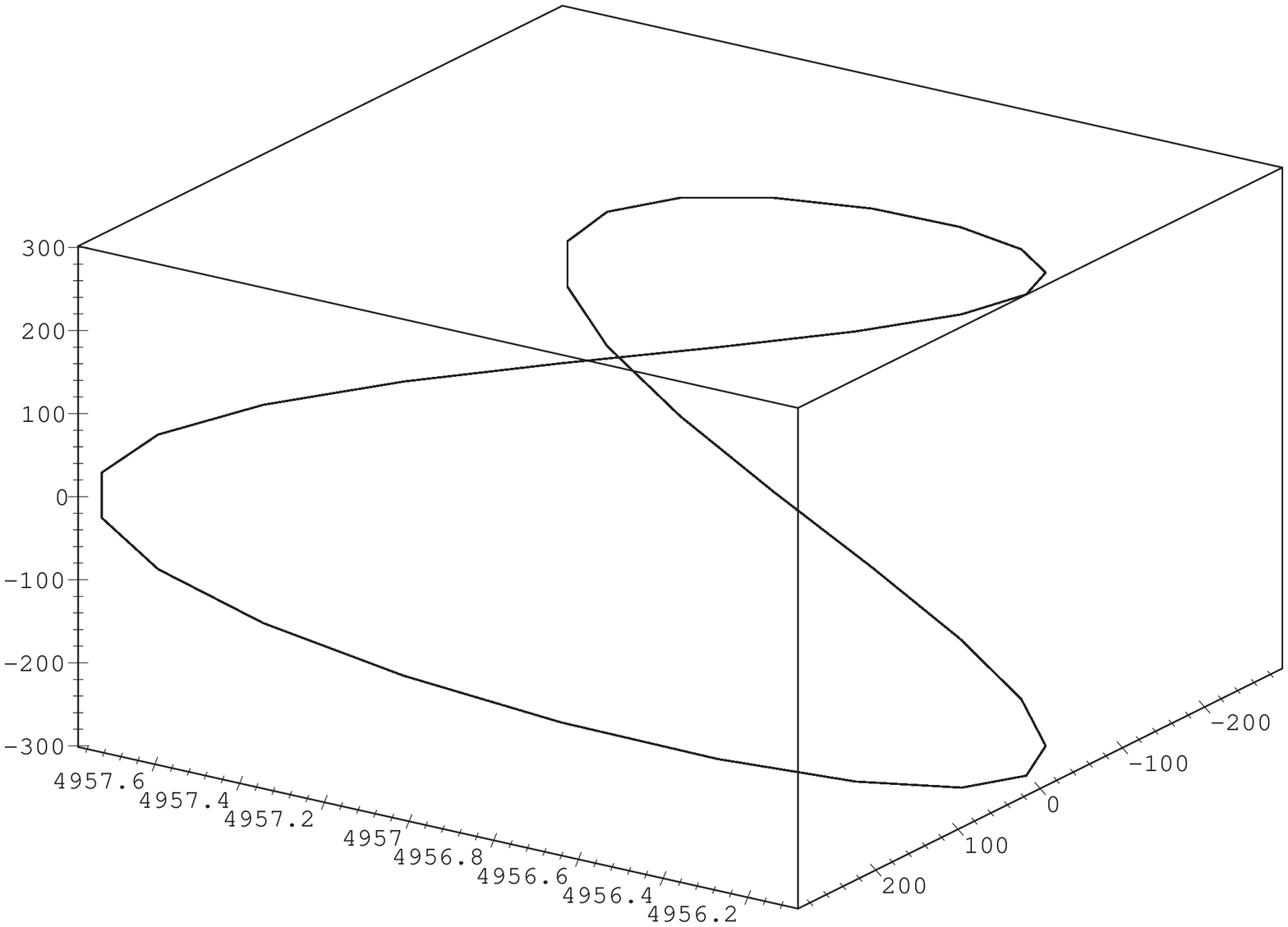}\\
\includegraphics[width=2.5in,angle=-90]{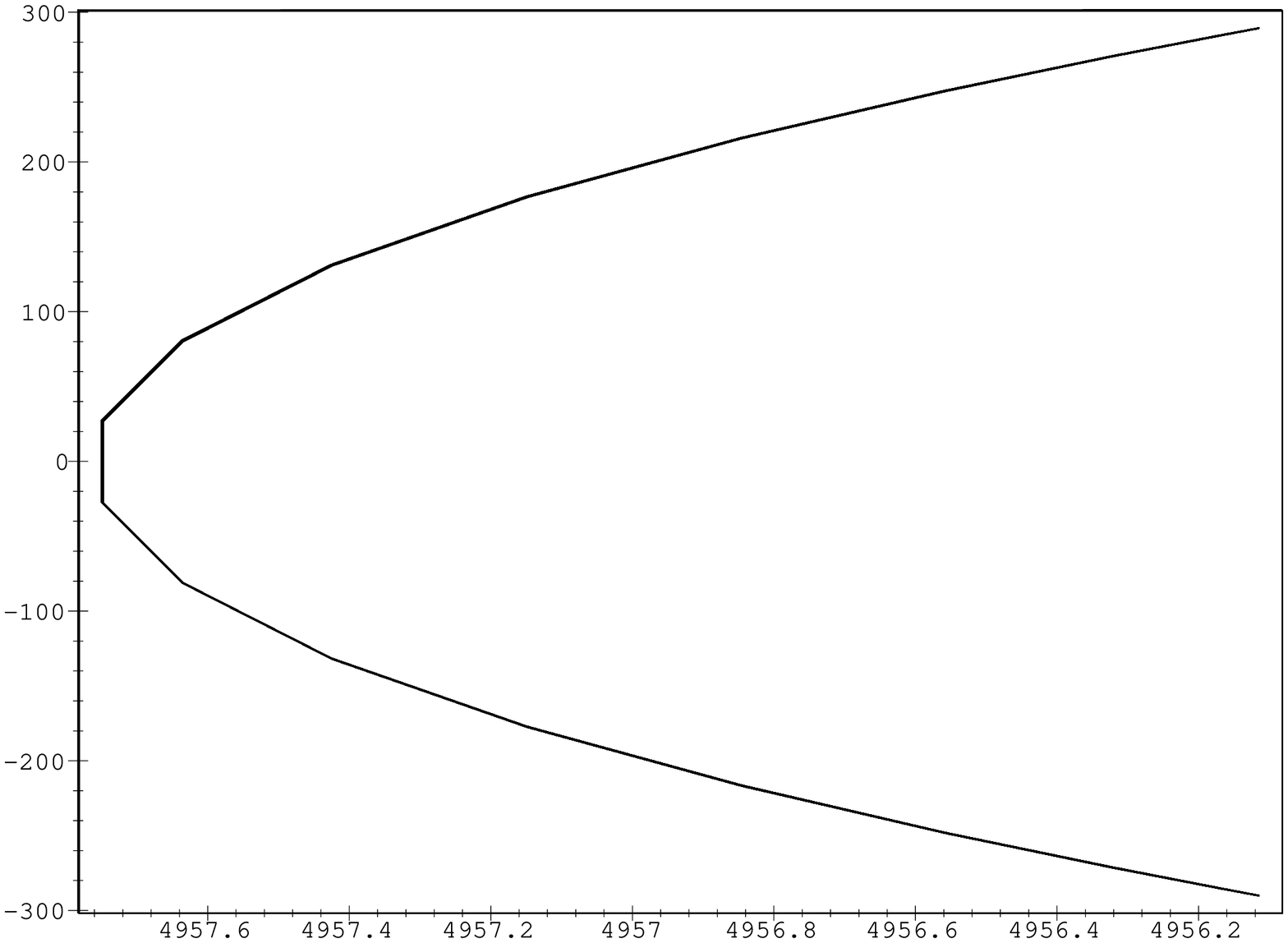}\\
\includegraphics[width=2.5in,angle=-90]{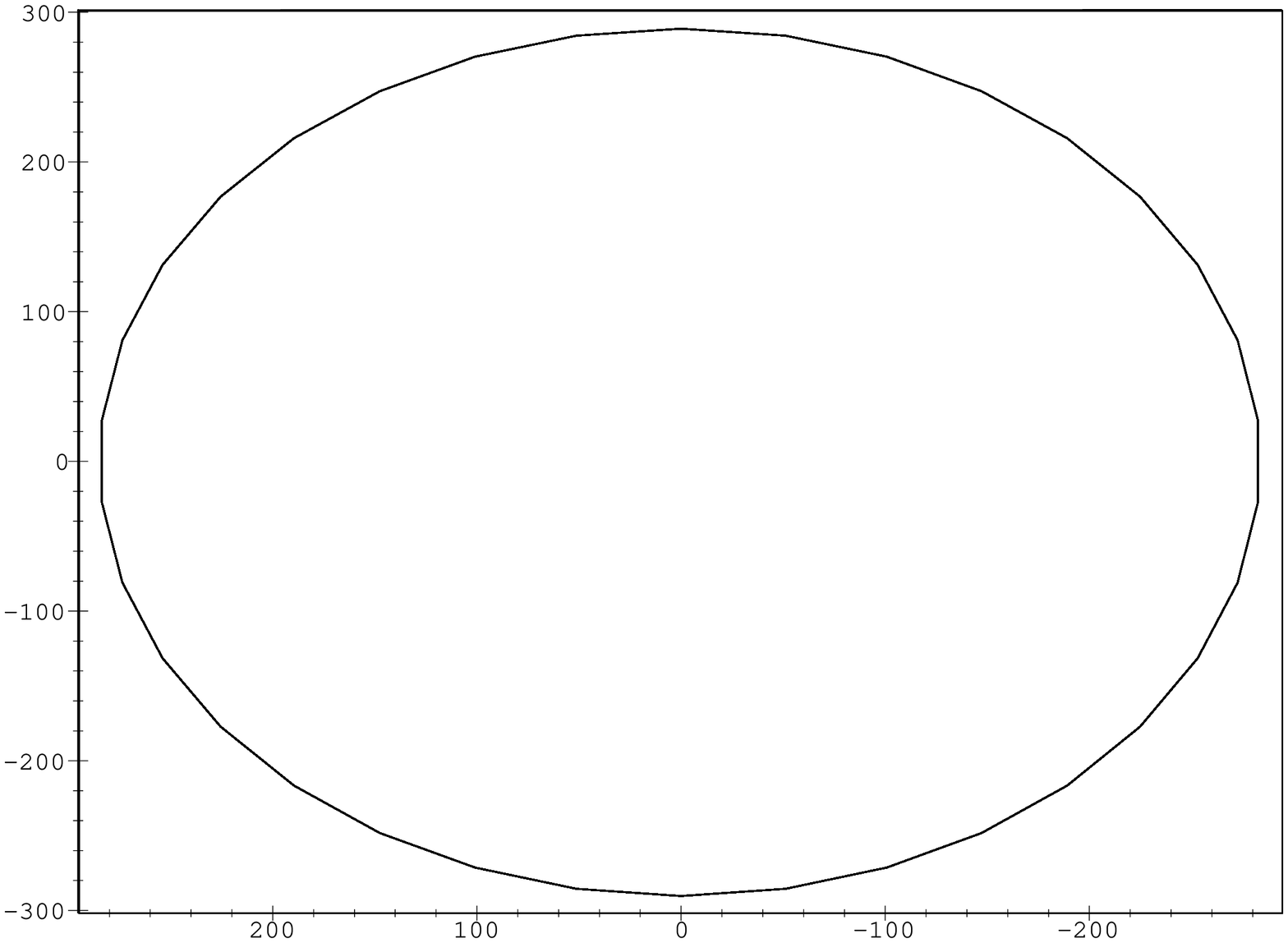}\\
\caption{\label{fig:goblet0rim} The boundary of the wavefront in
the critical case. Middle and bottom panels show side and front
views.}
\end{figure}

The second question is whether the rim of the goblet is planar,
namely, whether it lies on a constant-$z$  plane.  {\it The answer is
no.}   The rim of the goblet is a space curve, which is plotted
in Figure~\ref{fig:goblet0rim}.  It must be kept in mind that
the rim is the boundary of the wavefront.  This rim is removed
because it is the image of the point $r=0$ on the lens plane
under the wavefront lensing map, i.e.,
$\bw_{T,ep}^{sing} (0, \vartheta) = (\bs (0,\vartheta, T), z (0,\vartheta, T))$,
which is a space curve for fixed $T$.
Explicitly,  equations (\ref{eq-zwavefront}) and
(\ref{eq-swavefront})
yield that
\beqa
\label{eq-s-z-sep}
\bs (0,\vartheta, T) 
&=&
\frac{- \left(  T
       \  +  \   z_l\ |\hat{\bma}_{ep}^{sing} (\vartheta)|^2/2
        \right)
          \hat{\bma}_{ep}^{sing} (\vartheta)
         }{\displaystyle
           1
        \  + \  |\hat{\bma}_{ep}^{sing} (\vartheta)|^2/2 } \nonumber \\
  && \hspace{1.5in}  + \  z_l  \ \hat{\bma}_{ep}^{sing} (\vartheta),
\nonumber \\
&& \nonumber\\
z(0,\vartheta, T)
&=& 
\frac{T
                \  +  \  z_l\ |\hat{\bma}_{ep}^{sing} (\vartheta)|^2/2
                }{ 1 \  +  \  |\hat{\bma}_{ep}^{sing} (\vartheta)|^2/2}.
\nonumber \\
&&
\eeqa
As is seen from these equations,
the ultimate reason why the wavefronts have a whole elliptical curve's
worth of boundary points instead of only one point is that the
bending angle vector field $\hat{\bma}_{ep}^{sing}$ does not depend on $r$.

In rectangular
coordinates, the map $\hat{\bma}_{ep}^{sing}$ is ill-defined at the origin $\br=\bmo$.
It has a singularity of the type 0/0 since
$$
\hat{\bma}_{ep}^{sing} (\br) =
\frac{A_0^2\Big( (1- \epsilon)r_1, (1 + \epsilon)r_2\Big)}{
       \hp_{ep}^{sing} (\br)}.
$$
One can approach the singular point from different directions in
the lens plane and hope that the limit would be the same in all
directions. However, the limiting value of the
$\hat{\bma}_{ep}^{sing} (\br)$ as $\br\to \bmo$ is
direction-dependent, as signaled by the direction-dependent
bending angle $\vartheta$ in  (\ref{eq-s-z-sep}). In this
atypically singular case, obstructing one light ray (the one that
passes through the center of the lens) removes an elliptical
curve's worth of points on the wavefront. If a source is outside
the rim, then one lensed image is seen of the source, while two
images are seen if the source is inside the rim.  This phenomenon
seems to have been first noted on p. 188 of \cite{PLW}
for the case of a singular isothermal sphere (i.e., $r_c =
\epsilon =0$), where the rim is a circle, as discussed in
Subsection~\ref{subsec:singiso}

\subsubsection{Caustic Sheet}

The caustic sheet in this case is obtained by the same method
as in the case of a nonsingular elliptical potential.
The use of polar coordinates
 $(r, \vartheta)$ simplifies the calculation. We have
\beqan
\det\frac{\pa \bs_z}{\pa \br}
& = & 
    \cos(2\vartheta)\  \det   \frac{\pa \bs_z (\br)}{\pa (r,\vartheta)}
   \\ 
& = & 
    \cos(2\phi) \left(\frac{\partial x}{\partial r}
             \frac{\partial y}{\partial \vartheta}
            -\frac{\partial x}{\partial \vartheta}
             \frac{\partial y}{\partial r}
            \right),
\eeqan
where $x$ and $y$ are the cartesian components of the
lensing map $\bs_z$. In this case, the components reduce to
\beqa
\label{eq-sing-epxy}
x &=& \left(\frac{z}{z_l}\  r
         - \frac{(z-z_l) A_0(1+\epsilon)}
            {\sqrt{1+\epsilon\cos(2\vartheta)}}
        \right) \cos\vartheta, \nonumber \\
y &=& \left(\frac{z}{z_l}\  r
         - \frac{(z-z_l) A_0(1-\epsilon)}
            {\sqrt{1+\epsilon\cos(2\vartheta)}}
        \right) \sin\vartheta, 
\nonumber \\
\eeqa
and the vanishing of the Jacobian determinant
--- see equation (\ref{eq-det-dszdr}) ---
becomes
\beqan
0  &=&
\det \frac{\pa \bs_z}{\pa \br}
   =
    \frac{z\cos(2\vartheta)}{z_l(1\!+\!\epsilon\cos(2\vartheta))^{\frac32}}
    \Big[ z\Big((1\!+\!\epsilon\cos(2\vartheta))^{\frac32} \frac{r}{z_l}
\\
&& \hspace{1.2in}
     - \ A_0(1\!-\!\epsilon^2)\Big)
\  + \ z_l A_0(1\!-\!\epsilon^2)
       \Big].
\eeqan
The unphysical root $z=0$, which defines the plane of the
observer, is responsible for the
absence of a second component caustic sheet, the most notable aspect of
the caustic sheet in the singular case. The (one component) caustic
sheet is obtained from the non-trivial root, i.e.,
\begin{equation}\label{eq-sing-epzc}
    z=z_c (r, \vartheta)
\equiv -\frac{z_l A_0(1-\epsilon^2)}
         {(1+\epsilon\cos(2\vartheta))^{\frac32}\ r/z_l
         - A_0(1-\epsilon^2)}.
\end{equation}

\begin{figure}
\vbox{
\includegraphics[width=2.7in,angle=-90]{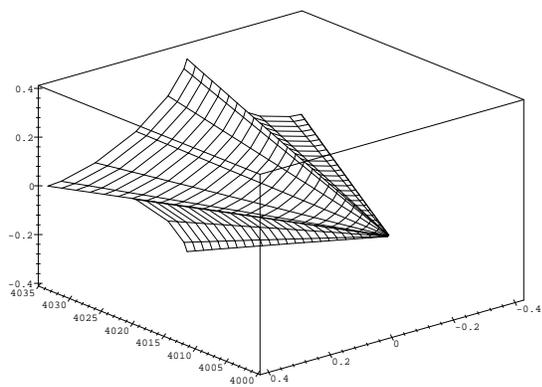}\\
\includegraphics[width=2.5in,angle=-90]{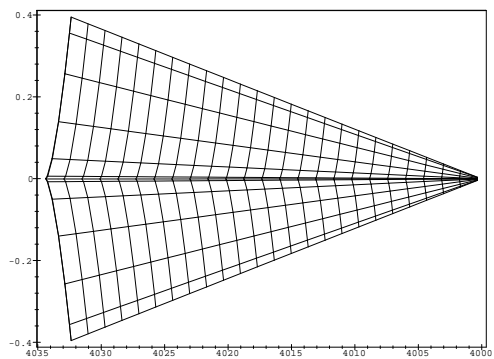}
        }
\caption{\label{fig:caustic0} The caustic sheet in the singular case}
\end{figure}

\begin{figure}
\includegraphics[width=2.5in,angle=-90]{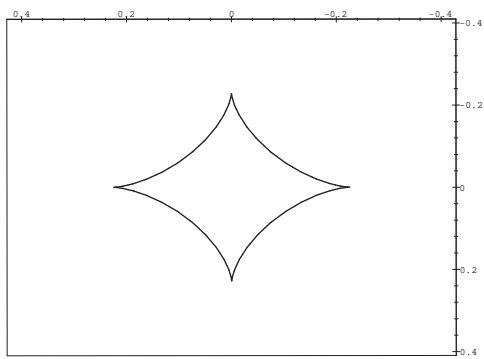}
\caption{\label{fig:planarcaustic0} Typical planar caustic in the
singular case.}
\end{figure}

A plot of the caustic sheet is shown in
Figure~\ref{fig:caustic0}. The surface has four cuspidal ridges
with a conical profile, which can clearly be appreciated in the
bottom panel of the figure. The tip of the caustic surface lies at
the center of the lens (origin) and is removed. Slicing the
caustic sheet with constant$-z$ planes we obtain the $z$-planar
caustic curves. The planar caustics are astroid-shaped curves with
four cusps, as can be seen in Figure~\ref{fig:planarcaustic0}.  A
source inside the caustic curve has four lensed images, while a
source outside has two.

\subsubsection{Caustic Surfing}

The singular elliptical potential provides an excellent
opportunity to illustrate the caustic surfing scheme. We start by
pointing out that the caustic sheet, which is parametrically given
by (\ref{eq-sing-epxy}) and (\ref{eq-sing-epzc}), can equivalently
be given by the following expression:
$$
S_0(\sfx) \equiv \left((1+\epsilon)x^2\right)^{\frac13}
+\left((1-\epsilon) y^2\right)^{\frac13}-\left(2\epsilon
A_0(z-z_l)\right)^{\frac23}=0
$$
This expression assumes that the apex of the lightcone where the
caustic lies is at the origin of coordinates, and the lens plane
is at a distance $z_l$ from it.

Suppose now that a source is moving on the light source plane with an
approximately constant velocity ${\sf V} = (V_1,V_2,0)$, carrying its
own lightcone with it. Then the caustic on the source's lightcone
moves as well, and at any given instant of conformal time $\sft$, assuming
the motion is sufficiently slow, the caustic surface can be given as 
$S(\sfx,\sft) =
S_0(\sfx-{\sf V}\sft)$.
Assume that
a space-borne telescope observes at time  $\sft =0$
an  image of the source at peak magnification. 
>From the location of the image, by the (comoving) lens map
we can determine the location $\bs$ of the source on the
source plane. But for our purposes we need to switch the point of
view and think of the lightcone of the source instead: The source
lies at the origin and the telescope at a location
$(x_0,y_0,z_0) = (-\bs \ell_{tl}/\ell_{sl},\ell_{tl}+\ell_{sl}),$
where $\ell_{tl}$ and $\ell_{sl}$ are the distance between the
telescope and the lens plane, and between the lens plane and the
source plane, resp.  Because $(x_0,y_0,z_0)$ lies on the
comoving future caustic sheet of the source at $\sft=0$, we have
\beq\label{eq-implicitt0}
S_0(\sfx_0)= 0 =
\left((1+\epsilon)x_0^2\right)^{\frac13}
+\left((1-\epsilon)
y_0^2\right)^{\frac13}-\left(2\epsilon
A_0\ell_{tl}\right)^{\frac23},
\eeq
where we have made the substitution $z_l = \ell_{tl}$.
Now, we
calculate the speed ${\sf v}_{min}$ that the telescope needs to
stay on the caustic sheet during the time increment $\delta \sft.$  
>From our discussion in
Section~\ref{sec-causticsurfing}, specifically
Eq.~(\ref{eq-vmin}), we have
\beqan
    {\sf v}_{min}
&=&
    \frac{ V_1\left[(1+\epsilon)/x_0\right]^\frac13
          +V_2\left[(1-\epsilon)/y_0\right]^\frac13}
          {\left[(1+\epsilon)/x_0\right]^\frac23
          +\left[(1-\epsilon)/y_0\right]^\frac23
          +\left[4\epsilon^2A_0^2/\ell_{tl}\right]^\frac23} \times
\\
&& \\
&&\hspace{0.3in}
    \left(\left(\frac{1+\epsilon}{x_0}\right)^\frac13,
          \left(\frac{1-\epsilon}{y_0}\right)^\frac13,
          -\left(\frac{4\epsilon^2A_0^2}{\ell_{tl}}\right)^\frac13
    \right)
\eeqan
In this expression, all the symbols are known in principle.  
The values of $x_0$ and $y_0$ can be  found
via the lens mapping at the observation event,
as explained above. Clearly, this may be a technically complex
procedure, considering that it is only a first step in a iterative
scheme to place the telescope on the caustic sheet.  Our main
purpose is to show that the calculation  is feasible in principle.

\subsection{Singular Isothermal Sphere Lens\label{subsec:singiso}}

\begin{figure}
\vbox{
\includegraphics[width=2.5in,angle=-90]{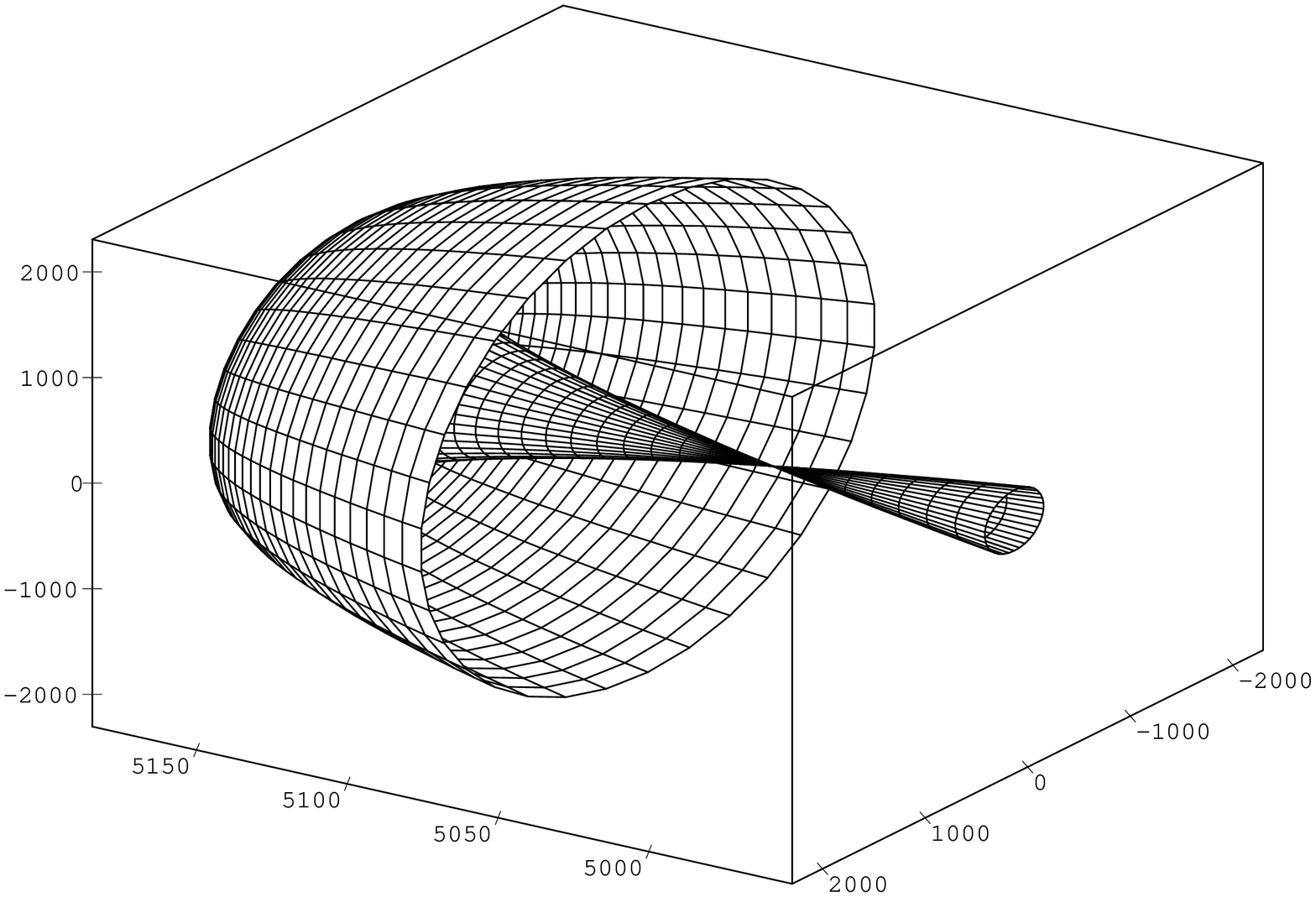}\\
\includegraphics[width=2.5in,angle=-90]{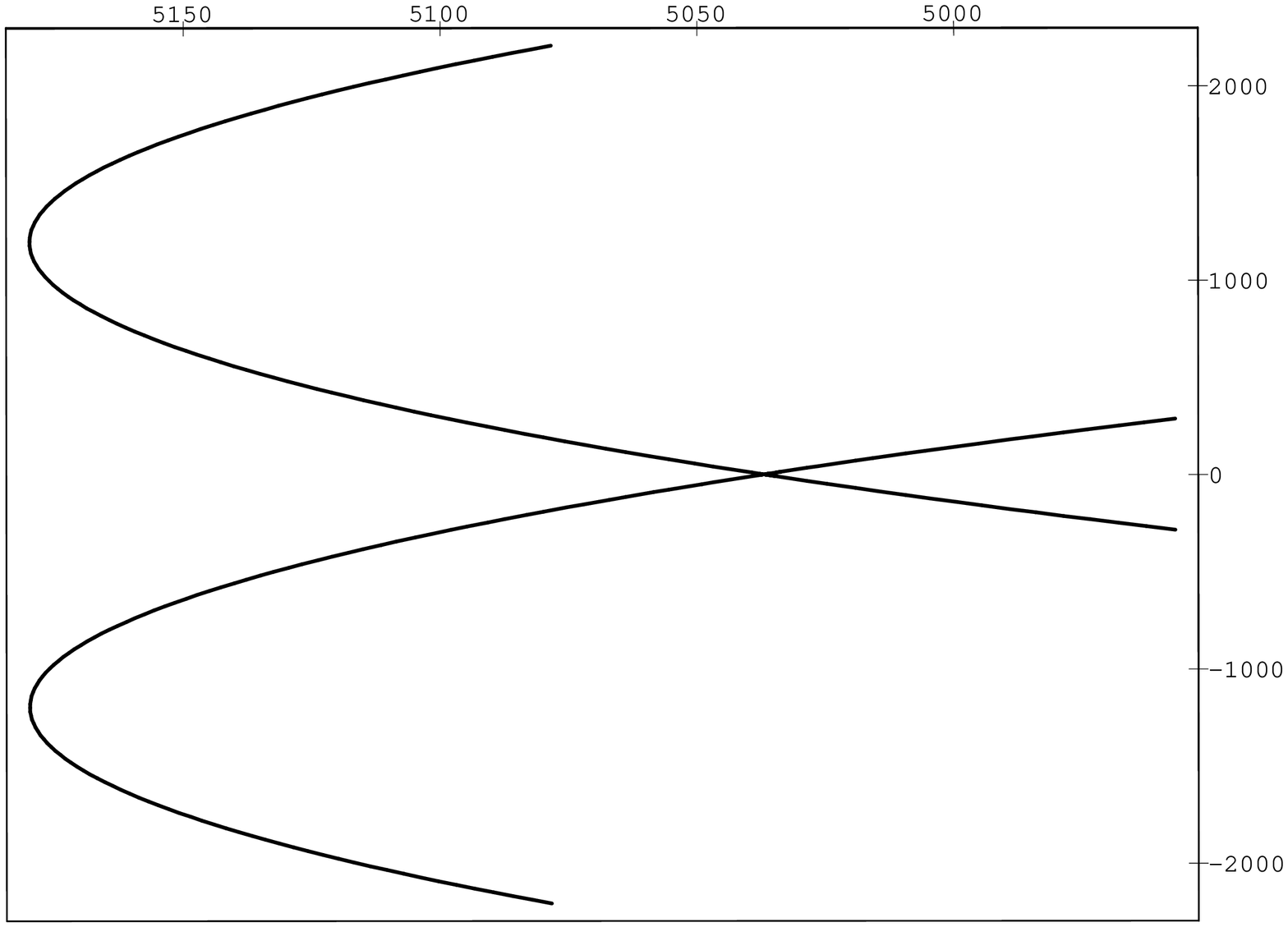}
      }
\caption{\label{fig:wavefront-sis} Wavefront due to a singular
isothermal sphere.  The top panel shows a field view, and the bottom panel
shows a slice or profile of the wavefront. Notice that the
wavefront is a surface of revolution around the optical axis.}
\end{figure}

For a singular isothermal sphere lens, we have $r_c = \epsilon
=0$.  The potential and bending angle vector are given
respectively as follows:
$$
    \hp_{ep}^{sis} (r,\vartheta)
    =  A_0 \  r,
\qquad
\hat{\bma}_{ep}^{sis} (r, \vartheta)
   =    A_0
    ( \cos\vartheta, \sin\vartheta).
$$
The wavefronts are similar to those of the
singular elliptical potential, except that
the throat of the goblet is a point and
the singular rim of the goblet is a circle
--- see Figure~\ref{fig:wavefront-sis}.
The caustic surface collapses to a line coinciding with
the portion of the optical axis beyond
the lens plane (i.e., $z>z_l$). Note the origin
on the lens plane, which is a singularity
of the lens, is not a part of the
caustic line.   We have that
each $z$-planar caustic is a point.

\section{Concluding remarks and outlook}

We have shown how to construct a lensing map that takes points on the
lens plane into points on a wavefront surface, as opposed to a source
plane. This represents a sort of ``instantaneous'' lens map, carrying
a sense of constant time. By contrast, the standard lens map carries
no sense of time at all. Additionally, the wavefront lensing map has
an  asymmetric Jacobian matrix. Notice that, barring multiple sheets,
the wavefront surface lies very close to a plane in the weak-field
case, as illustrated in most of our figures. In the figures the
optical axis is magnified several times in order to appreciate the
distance between the different sheets in the folded wavefront. Our
intial motivation for explicitly constructing a wavefront-lens
mapping was inspired by several works (e.g.,
\cite{BU80,Nt90,Ptt93,exactuniv,EFN01}).  
In this
respect,  it appears that an extension of our constructions in Secs.
III and IV beyond the weak field domain is
feasible~\cite{ehlers00,perlick}.

We took advantage of the conformally flat nature of the flat FL
universe in order to make use of a comoving lens equation, which in
addition to being comoving is also a conformal lens equation. A
conformally related flat spacetime exists, of course, in all three
types of FL universes, so in principle, our wavefront map could be
adapted for interpretation in the open or closed universes. However,
in such universes the conformal factor relating the FL universes to
the corresponding flat spacetime depends on the space point, as well
as the time, and the translation of our conformal present proper
length function  to a cosmological present proper length  is not at
all as direct as we found it in this work. Some of the subtleties
that would be involved in the translation in the open and closed
cases are treated in detail by Frittelli, Kling and Newman
\cite{FKN02}, where the conformally flat lens map and time delay are
transformed into the cosmological lensing map and time delay. 
Nonetheless,  we do not need to rely on the conformally related flat
space to calculate our present proper length function.

The individual wavefronts for the potentials illustrated proved
useful in visualizing the relationship between the location of the
source in reference to the caustic, and the number of images
observed. Additionally, the wavefronts of the singular potentials
helped explain the anomalous counting of images observed by  
Petters, Levine, and Wambsganss \cite{PLW}, p. 188.  The latter
showed that in the case of the singular potentials there is a simple
closed curve, that is not a caustic, but that separates regions in
the source plane where the number of images differs by one, rather
than two.  We have here shown that such a curve is the boundary of
the wavefront, it is not a caustic, and the number of images differs
by one less than in the regular case because one whole sheet of the
wavefront is missing due to an obstruction in the lens plane.

Lastly, we have taken a step towards a preliminary scheme  for
caustic surfing in a meaningful and consistent manner. The form of
the subsequent iterations remains an open problem, as does the
implementation of the procedure, particularly the measurement of the
transverse velocity of the the moving source that one intends to
follow.  Clearly the caustic surfing proposal depends only on the
caustic surface, and not on the method used to obtain it. But the
point is that optimal caustic surfing (with the least effort) is
achieved only by allowing the telescope to surf the caustic sheet,
rather than the planar caustics at fixed distance from the source. 
One could also image scenarios where a telescope may not ride a
caustic sheet, but move so as to stay inside certain chambers of the
caustic surface (compare with Gaudi and Gould \cite{GGa97}),
possibly near higher order singularities.  Future studies of the
aforementioned issues are clearly warranted.

\begin{acknowledgments}

We are indebetd to Scott Gaudi, Marek Kossowki, and Joachim
Wambsganss for discussions. We are especially thankful to J\"urgen
Ehlers for valuable critical feedback on the manuscript, in
particular regarding the difficulties with the validity of the
standard approximations in astrophysical gravitational lensing. S. F.
thanks Duke University where part of this work was carried out, and
gratefully  acknowledges support from NSF under grant PHY-0070624.
A.P. was supported by an Alfred P. Sloan Research Fellowship and NSF
Career grant DMS-98-96274. 

\end{acknowledgments}


\newpage

\mbox{}

\newpage

\end{document}